\documentclass[prd,aps,preprintnumbers,showpacs,amsfonts,floatfix,nofootinbib,superscriptaddress]{revtex4}
\usepackage{latexsym,graphicx,epsfig,psfrag,here}
\begin{document}

\preprint{DSF-2006/3 (Napoli)}

\title{Exclusive semileptonic and nonleptonic decays of the $B_c$ meson}

\author{Mikhail A. Ivanov}
\affiliation{Bogoliubov Laboratory of Theoretical Physics, \\
Joint Institute for Nuclear Research, 141980 Dubna, Russia}
\author{J\"{u}rgen G. K\"{o}rner}
\affiliation{Institut f\"{u}r Physik, Johannes Gutenberg-Universit\"{a}t, \\
D-55099 Mainz, Germany}
\author{Pietro Santorelli}
\affiliation{Dipartimento di Scienze Fisiche, Universit{\`a} di
Napoli "Federico II", \\
Istituto Nazionale di Fisica Nucleare, Sezione di Napoli, Italy}

\begin{abstract}
We study exclusive nonleptonic and semileptonic decays of the $B_c$-meson
within a relativistic constituent quark model previously developed by us.
For the nonleptonic decays we use the factorizing approximation.
We update our model parameters by using new
experimental data for the mass and the lifetime
of the $B_c$ meson and the leptonic decay
constants of the $D$-meson. We calculate the
branching ratios for a large set of exclusive nonleptonic and
semileptonic decays of the $B_c$ meson and compare our results with the
results of other studies.
As a guide for further experimental exploration we provide explicit formulas
for the full angular decay distributions in the cascade decays
$B_c^- \rightarrow J/\psi (\to l^+l^-) + \rho^-(\to \pi^- \pi^0)$ and
$B_c^- \to J/\psi (\to l^+l^-) + W^-_{\rm{off-shell}}
(\to l^- + \bar{\nu}_l)$.
\end{abstract}

\maketitle

\section{Introduction}
\label{sec:intro}

In 1998 the CDF Collaboration reported on the observation of the
bottom-charm $B_c$ meson at Fermilab \cite{CDF}
in the semileptonic decay mode  $B_c\to J/\psi +l +\nu$
with the $J/\psi$ decaying into muon pairs.
Values for the mass and the lifetime of the $B_c$ meson
were given as
 $M(B_c)=6.40\pm 0.39\pm 0.13$ GeV
and
 $\tau(B_c)=0.46^{+0.18}_{-0.16}({\rm stat})\pm 0.03({\rm syst})$ ps.
Recently, CDF reported first Run II  evidence for the $B_c$-meson
in the fully reconstructed
decay channel $B_c\to J/\psi+\pi$ with
$J/\psi\to\mu^+\mu^-$ \cite{Acosta:2005us}.
The mass value quoted for this decay channel is
$6.2857 \pm 0.0053({\rm stat.}) \pm 0.0012 ({\rm syst.})$ GeV
with errors significantly smaller than in the first
measurement. Also D0 has observed the $B_c$ in the semileptonic
mode $B_c\to J/\psi+\mu +X$ and reported preliminary evidence
that $M(B_c) = 5.95^{+0.14}_{-0.13} \pm 0.34$ GeV
and $\tau(B_c)= 0.45^{+0.12}_{-0.10} \pm 0.12$ ps \cite{Corcoran:2005ti}.

The $B_c$-meson is the lowest bound state of
two heavy quarks (charm and bottom) with open
flavor. The $B_c$-meson therefore decays weakly. It can decay via
(i) b-quark decay, (ii) c-quark decay, and (iii) the annihilation channel.
The modern state of art, starting from the pioneering paper \cite{{Lusignoli:1990ky}},
in the spectroscopy, production and decays of the $B_c$-meson can be found in the review
\cite{Brambilla:2004wf} and the published talk  \cite{Chang:2005ek}.

In this article we complete the analysis of
almost all accessible low-lying exclusive nonleptonic two-body
and semileptonic three-body  modes of the $B_c$-decays
within our relativistic constituent quark model
\cite{Ivanov:2000aj,Faessler:2002ut,Ivanov:2002un,Ivanov:2005fd}.
We update the model parameters by using the latest
experimental data on the $B_c$-mass \cite{Acosta:2005us}
and the weak decay constant $f_D$ \cite{Artuso:2005ym}.
We give a set of numerical values
for the leptonic, semileptonic and nonleptonic partial decay widths
of the $B_c$-meson and compare them with the results of other approaches.
We provide explicit formulas for
the angular decay distributions of the cascade decays
$B_c^- \rightarrow J/\psi (\to l^+l^-) + \rho^-(\to \pi^- \pi^0)$
and $B_c^- \to J/\psi (\to l^+l^-) + W^-_{\rm{off-shell}}
(\to l^- + \bar{\nu}_l)$ by using the methods described in \cite{ks90} and
subsequently applied to various other cascade decay processes (see
\cite{Korner:1990yx,Korner:1992kj,bkkz93,Faessler:2002ut,Kadeer:2005aq}).
For the nonleptonic decay $B_c^- \rightarrow J/\psi  + \rho^-$ we also include
lepton mass and $T$--odd effects in our analysis. These angular
decay distributions may be of help in analyzing the cascade decay data. Also,
by analyzing the cascade angular decay distributions, one can learn more
details about the spin dynamics of the decay process than from the rate
analysis alone.

\section{Model}
\label{sec:model}

The coupling of a meson $H(q_1\bar q_2)$ to its constituent
quarks $q_1$ and $\bar q_2 $ is described  by the Lagrangian \cite{QCM,RCQM}
\begin{equation}
\label{Lagr_str}
{\mathcal L}_{\rm int\, Hqq}(x) = g_H H(x)\int\!\! dx_1 \!\!\int\!\!
dx_2 F_H (x,x_1,x_2)\bar q_2(x_2)\Gamma_Hq_1(x_1) \, + {\rm h.c.}
\end{equation}
Here, $\Gamma_H$ is a Dirac matrix or a string of Dirac matrices
which projects onto the spin quantum number of the meson field $H(x)$.
The function $F_H$ is related to the scalar part of the
Bethe-Salpeter amplitude and characterizes the finite size of the
meson. To satisfy translational invariance the function $F_H$
has to fulfil the identity $F_H(x+a,x_1+a,x_2+a)=F_H(x,x_1,x_2)$ for
any four-vector a. In the following we use
a specific  form for the scalar vertex function
\begin{equation}
\label{vertex}
F_H(x,x_1,x_2)=\delta(x - c^1_{12} x_1 - c^2_{12} x_2) \Phi_H((x_1-x_2)^2)
\end{equation}
where $\Phi_H$ is the correlation function of the two constituent quarks
with masses $m_{q_1}$, $m_{q_2}$ and the mass ratios
$c^{i}_{ij}=m_{q_i}/(m_{q_i}+m_{q_j})$.

The coupling constant $g_H$ in Eq.~(\ref{Lagr_str}) is determined by the
so-called {\it compositeness condition} originally proposed in~\cite{Z=0},
and extensively used in
\cite{QCM,RCQM,Ivanov:2000aj,Faessler:2002ut,Ivanov:2005fd}.
The compositeness condition requires that the renormalization constant of
the elementary meson field $H(x)$ is set to zero
\begin{equation}
\label{z=0}
Z_H \, = \, 1 - \, \frac{3g^2_H}{4\pi^2} \,
\tilde\Pi'_H(m^2_H) \, = \, 0
\end{equation}
where $\tilde\Pi^\prime_H$ is the derivative of the meson mass operator.
To clarify the physical meaning of the compositeness  condition in Eq.~(\ref{z=0}),
we first want to remind the reader
that the renormalization constant $Z_H^{1/2}$ can also interpreted as
the matrix element between the physical and the corresponding bare state.
The condition  $Z_H=0$ implies that the physical state does not contain
the bare state and is appropriately described as a bound state.
The interaction Lagrangian of Eq.~(\ref{Lagr_str}) and
the corresponding free parts of the Lagrangian describe
both the constituents (quarks) and the physical particles (hadrons)
which are viewed as the bound states of the quarks.
As a result of the interaction, the physical particle is dressed,
i.e. its mass and wave function have to be renormalized.
The condition $Z_H=0$ also effectively excludes
the constituent degrees of freedom from the space of physical states.
It thereby guarantees that there is no double counting
for the physical observable under consideration.
The constituents exist only in  virtual states.
One of the corollaries of the compositeness condition is the absence
of a direct interaction of the dressed charged particle with the
electromagnetic field. Taking into account both the tree-level
diagram and the diagrams with the self-energy insertions into the
external legs (i.e. the tree-level diagram times $Z_H -1$) yields
a common factor $Z_H$  which is equal to zero. We refer the interested
reader to our previous papers
\cite{QCM,RCQM,Ivanov:2000aj,Faessler:2002ut,Ivanov:2005fd}
where these points are discussed in more detail.

In the case of the pseudoscalar and vector mesons
the derivative of the meson mass operator appearing in Eq.~(\ref{z=0})
can be calculated in the following way:

\begin{eqnarray}\label{Mass-operator}
&&
\tilde\Pi'_P(p^2) \, =
\frac{p^\alpha}{2p^2}\,
\frac{d}{dp^\alpha}\,
\int\!\! \frac{d^4k}{4\pi^2i} \tilde\Phi^2_P(-k^2)
{\rm tr} \biggl[\gamma^5 \tilde S_1(k+c^1_{12} p)
\gamma^5 \tilde S_2(k-c^2_{12} p) \biggr] \, ,
\nonumber\\
&&\nonumber\\
&&
 \tilde\Pi'_V(p^2) \, =
\frac{1}{3}\left( g_{\mu\nu}-\frac{p_\mu p_\nu}{p^2}\right)
\frac{p^\alpha}{2p^2}\,
\frac{d}{dp^\alpha}\,
\int\!\! \frac{d^4k}{4\pi^2i} \tilde\Phi^2_V(-k^2)
{\rm tr} \biggl[\gamma^\mu\tilde S_1(k+c^1_{12} p)
\gamma^\nu\tilde S_2(k-c^2_{12} p) \biggr].
\end{eqnarray}
where $\Phi_H((x_1-x_2)^2)$ is a correlation function
and $\tilde S_i(k)$ is the quark propagator.
We have used free fermion propagators for the valence quarks
given by
\begin{equation}
\label{quark-prop}
\tilde S_i(k)=\frac{1}{m_{q_i}-\not\! k}
\end{equation}
with an effective constituent quark mass $m_i$. As discussed
in~\cite{RCQM} we have assumed that the meson mass $m_H$ lies
below the constituent quark threshold, i.e. we have
\begin{equation}
\label{conf}
m_H < m_{q_1} + m_{q_2}.
\end{equation}

Since the transitions in our approach are described by one-loop
quark diagrams the condition (\ref{conf}) guarantees that
there are no imaginary parts in our  physical transition amplitudes.
For the constituent quark masses that we use
this is satisfied for the  low-lying pseudoscalar mesons
$\pi$, $K$, $D$, $D_s$, $B$, $B_s$, $B_c$ and
$\eta_c$ and also for the $J/\psi$ but is no longer true
for the light vector mesons ($\rho$, $K^\ast$),
the heavy flavored vector mesons ($D^\ast$ and $B^\ast$)
and for the $p$--wave and excited charmonium states considered in this paper.
We have therefore employed
\cite{Ivanov:2000aj,Faessler:2002ut}
identical masses for all
heavy pseudoscalar and vector flavored mesons
($m_{B^\ast}=m_{B}$, $m_{D^\ast}=m_{D}$)
and for all $p$--wave and excited charmonium states
\cite{Ivanov:2005fd}
in our matrix element calculations
but have used physical masses in the phase space calculation.
This is quite a reliable approximation
for the heavy mesons because the corresponding  mass splittings
are relatively small. For the light vector mesons ($\rho$, $K^\ast$) this
approximation is not very good. However, in the present application
the light vector mesons do not explicitly enter into the
decay dynamics described by the transition matrix elements. They
contribute only in the form of the leptonic decay constants $f_\rho=210$ MeV
and $f_{K^\ast}=217$ MeV for which we use the experimental values.
We emphasize that the quark mass function appearing in
the Dyson-Schwinger-Equations (DSE) studies \cite{DSE-heavy}
is almost constant in the case of the $b$-quark. This is true
to a lesser extent for the $c$-quark. However, in the case of light
$u,d$ and $s$ quarks the momentum-dependent dressing is essential.

\section{Nonleptonic decays of the $B_c$-meson}
\label{sec:nonlep}

The effective Hamiltonian describing the $B_c$-nonleptonic
decays is given by
\begin{eqnarray}
{\mathcal H}_{\rm eff} &=&
-\frac{G_F}{\sqrt{2}}\,
\left\{
V_{cb} V^\dagger_{ud}
\left[
c_1\, (\bar cb)_{V-A}(\bar du)_{V-A}+c_2\, (\bar db)_{V-A}(\bar cu)_{V-A}
\right]
\right.
\nonumber\\
&&\hspace*{1cm}
+\,V_{cb} V^\dagger_{us}
\left[
c_1\, (\bar cb)_{V-A}(\bar su)_{V-A}+c_2\, (\bar sb)_{V-A}(\bar cu)_{V-A}
\right]
\nonumber\\
&&\hspace*{1cm}
+\,V_{cb} V^\dagger_{cd}
\left[
c_1\, (\bar cb)_{V-A}(\bar dc)_{V-A}+c_2\, (\bar db)_{V-A} (\bar cc)_{V-A}
\right]
\nonumber\\
&&\hspace*{1cm}
+\,V_{cb} V^\dagger_{cs}
\left[
c_1\, (\bar cb)_{V-A}(\bar sc)_{V-A}+c_2\,(\bar sb)_{V-A} (\bar cc)_{V-A}
\right]
\nonumber\\
&&\nonumber\\
&&\hspace*{1cm}
+\,V_{ub} V^\dagger_{ud}
\left[
c_1\, (\bar ub)_{V-A}(\bar du)_{V-A}+c_2\,(\bar db)_{V-A}  (\bar uu)_{V-A}
\right]
\nonumber\\
&&\hspace*{1cm}
+\,V_{ub} V^\dagger_{us}
\left[
c_1\, (\bar ub)_{V-A}(\bar su)_{V-A}+c_2\,(\bar sb)_{V-A}  (\bar uu)_{V-A}
\right]
\nonumber\\
&&\hspace*{1cm}
+\,V_{ub} V^\dagger_{cd}
\left[
c_1\, (\bar ub)_{V-A}(\bar dc)_{V-A}+c_2\,(\bar db)_{V-A}  (\bar uc)_{V-A}
\right]
\nonumber\\
&&\hspace*{1cm}
+\,V_{ub} V^\dagger_{cs}
\left[
c_1\, (\bar ub)_{V-A}(\bar sc)_{V-A}+c_2\,(\bar sb)_{V-A}  (\bar uc)_{V-A}
\right]
\nonumber\\
&&\nonumber\\
&&\hspace*{1cm}
+\,V_{cs} V^\dagger_{ud}
\left[
c_1\, (\bar cs)_{V-A}(\bar du)_{V-A}+c_2\,(\bar cu)_{V-A}  (\bar ds)_{V-A}
\right]
\nonumber\\
&&\hspace*{1cm}
+\,V_{cs} V^\dagger_{us}
\left[
c_1\, (\bar cs)_{V-A}(\bar su)_{V-A}+c_2\,(\bar cu)_{V-A}  (\bar ss)_{V-A}
\right]
\nonumber\\
&&\nonumber\\
&&\hspace*{1cm}
+\,V_{cd} V^\dagger_{ud}
\left[
c_1\, (\bar cd)_{V-A}(\bar du)_{V-A}+c_2\,(\bar cu)_{V-A}  (\bar dd)_{V-A}
\right]
\nonumber\\
&&\hspace*{1cm}
+\,V_{cd} V^\dagger_{us}
\left[
c_1\, (\bar cd)_{V-A}(\bar su)_{V-A}+c_2\,(\bar cu)_{V-A}  (\bar sd)_{V-A}
\right]
\left.
\right\} + {\rm h.c.}\,,
\label{hamilt}
\end{eqnarray}
where the subscript $V-A$ refers to the usual left--chiral current
$O^\mu=\gamma^\mu(1-\gamma^5)$.
We calculate the nonleptonic $B_c$-decay widths by
using {\it naive} factorization. First, we give
the necessary definitions of the leptonic decay constants,
invariant form factors and helicity amplitudes as they were introduced
in our paper \cite{Ivanov:2005fd}.

The leptonic decay constants are defined by
\begin{eqnarray}
&&
M(H_{12}\to \bar l \nu)=
\frac{G_F}{\sqrt{2}}\,V_{q_1q_2}\,{\mathcal M}_H^\mu(p)\,
\bar u_l(k_l)\,O^\mu\,u_\nu(k_\nu),
\nonumber\\
&&\nonumber\\
&&
{\mathcal M}_H^\mu(p) = -\,3\,g_{12}\,
\int\!\frac{d^4k}{(2\,\pi)^4\,i}\,
\widetilde\Phi_{12}\left(-k^2\right)\,
{\rm tr}\left[\Gamma_H\,\widetilde S_2(k-c^2_{12}\, p)\,
O^\mu\,\widetilde S_1(k+c^1_{12}\, p)\,\right]\,,
\nonumber\\
&&
\Gamma_P=i\,\gamma^5,\,\,\, \Gamma_V=\varepsilon_V\cdot \gamma,
\nonumber\\
&&\nonumber\\
&&
{\mathcal M}_P^\mu(p)= -i f_P p^\mu,
\hspace{1cm}
{\mathcal M}_V^\mu(p)= f_V m_V \varepsilon_V^\mu.
\label{lept}
\end{eqnarray}

The semileptonic  decays of the $B_c$-meson may be
induced by either a b-quark or a c-quark transition.
For the sake of brevity, we use a notation where $q_1\equiv b$ and
$q_3\equiv c$ whereas $q_2$ denotes either of $c,u,d,s$.
\begin{eqnarray}
M(H_{13}\to H_{23}+\bar l\nu) & = &
\frac{G_F}{\sqrt{2}}\,V_{q_1q_2}\,{\mathcal M^\mu_{12}}(p_1,p_2)\,
\bar u_l(k_l)\,O^\mu\,u_\nu(k_\nu),
\hspace{0.5cm} {\rm b-decay},
\nonumber\\
M(H_{13}\to H_{12}+\bar l\nu) & = &
\frac{G_F}{\sqrt{2}}\,V_{q_2q_3}\,{\mathcal M^\mu_{23}}(p_1,p_2)\,
\bar u_l(k_l)\,O^\mu\,u_\nu(k_\nu),
\hspace{0.5cm} {\rm c-decay},
\nonumber\\
&&\nonumber\\
{\mathcal M^\mu_{12}} & = &
-\,3\,g_{13}\,g_{23}\,
\int\!\frac{d^4k}{(2\,\pi)^4\,i}\,
\widetilde\Phi_{13}\left(-(k+c^3_{13}\,p_1)^2\right)\,
\widetilde\Phi_{23}\left(-(k+c^3_{23}\,p_2)^2\right)\,
\nonumber\\
&&
\times{\rm tr}\left[\,i\,\gamma^5\,\widetilde S_3(k)\,
\Gamma_{32}\,\widetilde S_2(k+p_2)\,O^\mu\,\widetilde S_1(k+p_1)\,\right],
\\
&&\nonumber\\
{\mathcal M^\mu_{23}}  & = &
-\,3\,g_{13}\,g_{12}\,
\int\!\frac{d^4k}{(2\,\pi)^4\,i}\,
\widetilde\Phi_{13}\left(-(k-c^1_{13}\,p_1)^2\right)\,
\widetilde\Phi_{12}\left(-(k-c^1_{12}\,p_2)^2\right)\,
\nonumber\\
&&
\times{\rm tr}\left[\,i\,\gamma^5\,\widetilde S_3(k-p_1)\,
O^\mu\,\widetilde S_2(k-p_2)\Gamma_{21}\,\widetilde S_1(k)\,\right].
\end{eqnarray}

We mention that we have checked in \cite{Ivanov:2000aj} that, in the heavy
mass limit, our form factors satisfy the HQET relations
written down in \cite{Jenkins:1992nb}.

The invariant form factors for the semileptonic $B_c$-decay
into the hadron with spin $S=0,1,2$ are defined by

\begin{eqnarray}
{\mathcal M^\mu_{\,S=0}} &=&
P^\mu\,F_+(q^2)+q^\mu\,F_-(q^2),\label{ff0}\\
&&\nonumber\\
{\mathcal M^\mu_{\,S=1}}&=&
\frac{1}{m_1+m_2}\,\epsilon^\dagger_\nu\,
\left\{\,
-\,g^{\mu\nu}\,Pq\,A_0(q^2)+P^\mu\,P^\nu\,A_+(q^2)
+q^\mu\,P^\nu\,A_-(q^2)
+i\,\varepsilon^{\mu\nu\alpha\beta}\,P_\alpha\,q_\beta\,V(q^2)\right\},
\label{ff1}\\
&&\nonumber\\
{\mathcal M^\mu_{\,S=2}}&=&
\epsilon^\dagger_{\nu\alpha}\,
\left\{\,
g^{\mu\alpha}\,P^\nu\,T_1(q^2)
+P^\nu\,P^\alpha\,\left[\,P^\mu\,T_2(q^2)+q^\mu\,T_3(q^2)\,\right]
+i\,\varepsilon^{\mu\nu\delta\beta}\,P^\alpha\,P_\delta\,q_\beta\,T_4(q^2)
\right\},\label{ff2}\\
&&\nonumber\\
P &=&p_1+p_2, \qquad q=p_1-p_2.
\nonumber
\end{eqnarray}
The form factor expansions cover both the $J^P \to (J^P)\,'$ cases
$0^- \to (0^-,1^-,2^-)$ and $0^- \to (0^+,1^+,2^+)$ needed in this paper.

One has to note that the form factors for the c-decay
can be obtained from the form factors for the b-decay
by simply exchanging the bottom and charm masses:

\begin{eqnarray}
&&\hspace*{-1cm}
 F_\pm^{\rm c-decay}(m_{q_1},m_{q_2},m_{q_3};q^2) =
-F_\pm^{\rm b-decay}(m_{q_3},m_{q_2},m_{q_1};q^2),
\\
&&\hspace*{-1cm}
 A_i^{\rm c-decay}(m_{q_1},m_{q_2},m_{q_3};q^2) =
-A_i^{\rm b-decay}(m_{q_3},m_{q_2},m_{q_1};q^2), \quad (i=0,\pm),
\\
&&\hspace*{-1cm}
V^{\rm c-decay}(m_{q_1},m_{q_2},m_{q_3};q^2) =
V^{\rm b-decay}(m_{q_3},m_{q_2},m_{q_1};q^2),
\end{eqnarray}
where we omit the explicit dependence on the ingoing
and outgoing masses and size parameters.

It is convenient to express all physical observables
through the helicity form factors $H_m$.
The helicity form factors $H_m$ can be expressed in terms of
the invariant form factors in the following way \cite{Ivanov:2000aj}:

\vspace{0.5cm}
\noindent
(a) Spin $S=0$:

\begin{eqnarray}
H_t &=& \frac{1}{\sqrt{q^2}}
\left\{(m_1^2-m_2^2)\, F_+ + q^2\, F_- \right\}\,,
\nonumber\\
H_\pm &=& 0\,,
\label{helS0b}\\
H_0 &=& \frac{2\,m_1\,|{\bf p_2}|}{\sqrt{q^2}} \,F_+ \,.
\nonumber
\end{eqnarray}

\vspace{0.5cm}
\noindent
(b) Spin $S=1$:

\begin{eqnarray}
H_t &=&
\frac{1}{m_1+m_2}\frac{m_1\,|{\bf p_2}|}{m_2\sqrt{q^2}}
\left\{ (m_1^2-m_2^2)\,(A_+ - A_0)+q^2 A_- \right\},
\nonumber\\
H_\pm &=&
\frac{1}{m_1+m_2}\left\{- (m_1^2-m_2^2)\, A_0
\pm 2\,m_1\,|{\bf p_2}|\, V \right\},
\label{helS1c}\\
H_0 &=&
\frac{1}{m_1+m_2}\frac{1}{2\,m_2\sqrt{q^2}}
\left\{-(m_1^2-m_2^2) \,(m_1^2-m_2^2-q^2)\, A_0
+4\,m_1^2\,|{\bf p_2}|^2\, A_+\right\}.
\nonumber
\end{eqnarray}

\vspace{0.5cm}
\noindent
(c) Spin $S=2$:
\begin{eqnarray}
H_t &=&
\sqrt\frac{2}{3}\,\frac{m_1^2\,  |{\bf p_2}|^2}{m_2^2\,\sqrt{q^2}}
\,\left\{T_1+\left[ |{\bf p_2}|^2+E_2\, q_0+m_1\, q_0\right]\,T_2
+q^2\, T_3\right\},
\nonumber\\
H_\pm &=&
\sqrt{\frac{1}{2}}
\frac{m_1\,|{\bf p_2}|}{m_2}\,
\left\{T_1\pm 2\, m_1|{\bf p_2}|\,T_4 \right\},
\label{helS2c}\\
H_0 &=&
\sqrt\frac{1}{6}\,
\frac{m_1\,|{\bf p_2}|}{m_2^2\,\sqrt{q^2}}
\left\{ (m_1^2-m_2^2-q^2)\, T_1 + 4\,m_1^2\,|{\bf p_2}|^2\,T_2 \right\},
\nonumber
\end{eqnarray}
where $|{\bf p_2}|=\lambda^{1/2}(m_1^2,m_2^2,q^2)/(2\,m_1)$,
$E_2 = (m_1^2+m_2^2-q^2)/(2\,m_1)$  and
$q_0=  (m_1^2-m_2^2+q^2)/(2\,m_1)$
are the momentum and energies of the outgoing particles
in the $B_c$ rest frame.
The widths of the semileptonic and nonleptonic decays of the $B_c$-meson
can be conveniently expressed in terms of the helicity form factors
\footnote{ As regards the transverse helicity amplitudes $H_\pm$
in Eqs.~(\ref{helS1c}) and (\ref{helS2c}) we have corrected
a sign error in our previous paper \cite{Ivanov:2005fd}.}.
The relevant width formulas are given in the Appendix.

\section{Numerical results}

In this paper we update the model parameters by using
the new values for the $B_c$-mass reported by
the CDF Coll. \cite{Acosta:2005us} the new value of the leptonic decay
constant$f_D$ reported by the CLEO Coll. \cite{Artuso:2005ym}
and lattice simulations
\cite{Aubin:2005ar,Chiu:2005ue,Gray:2005ad,Wingate:2003gm,
Lellouch:2000tw,Becirevic:1998ua}. The updated values
of the quark masses and size parameters are given by
Eq.~(\ref{fitmas}) and Eq.~(\ref{fitsize}), respectively.

\begin{equation}
\def\arraystretch{2}
\begin{array}{ccccc}
     m_u        &      m_s        &      m_c       &     m_b        &   \\
\hline
 \ \ 0.223\ \   &  \ \ 0.344\ \   &  \ \ 1.71\ \   &  \ \ 5.09\ \   &
 \ {\rm GeV} \\
\end{array}
\label{fitmas}
\end{equation}

\begin{equation}
\def\arraystretch{2}
\begin{array}{ccccccccccc}
 \Lambda_\pi   & \Lambda_K   & \Lambda_D        &  \Lambda_{D^\ast} &
 \Lambda_{D_s} & \Lambda_{B} & \Lambda_{B^\ast} &  \Lambda_{B_s}    &
 \Lambda_{B_c} & \Lambda_{cc}&  \\
\hline
\ \ 1.08 \ \  & \ \ 1.60 \ \ & \ \ 2.01\ \      & \ \ 1.46 \ \      &
\ \ 2.01 \ \  & \ \ 2.14 \ \ & \ \ 1.90\ \      & \ \ 2.14 \ \      &
\ \ 2.14 \ \  & \ \ 2.53 \ \ &  \  {\rm GeV} \\
\end{array}
\label{fitsize}
\end{equation}

\noindent The quality of the fit may be assessed from the entries
in Table~\ref{tab:leptonic}.

The calculation of the semileptonic and nonleptonic
decay widths is straightforward. For the CKM-matrix elements we use
\begin{equation}
\def\arraystretch{2}
\begin{array}{ccccccc}
|V_{ud}|     & |V_{us}|      & |V_{cd}|    & |V_{cs}|     & |V_{cb}|       &
|V_{ub}| \\
\hline
\ \ 0.975 \ \ & \ \ 0.224 \ \  & \ \ 0.224\ \ & \ \ 0.974 \ \  & \ \ 0.0413 \ \ &
\ \ 0.0037 \\
\end{array}
\label{CKM}
\end{equation}

The results of our evaluation of the branching ratios of the exclusive
semileptonic and nonleptonic $B_c$ decays appear in
Tables~\ref{tab:Bc-semlep1}$\div$\ref{tab:Bc-nonlep2}.

In the presentation of our results we
shall closely follow the format of the reviews of Kiselev
\cite{Kiselev:2003mp,Brambilla:2004wf}. Table~\ref{tab:Bc-semlep1} contains
our predictions for the exclusive semileptonic $B_c$ decays into ground state
charmonium states, and into ground state charm and bottom meson states.
Table~\ref{tab:Bc-semlep2} contains our predictions for the exclusive
semileptonic $B_c^-$ decays into $p$--wave charmonium states, and into the
orbital excitation of the charmonium state $\psi(3836)$. In
Table~\ref{tab:Bc-nonlepa1a2} we list our predictions for exclusive
nonleptonic decay widths of the $B_c$ meson using the factorization
hypothesis. In order to facilitate a comparison with other dynamical models
we list our results for general values of the effective Wilson coefficients
of the operator product expansion $a_1$ and $a_2$.

We then specify the values
of the effective Wilson coefficients. We take
$a_1^c=1.20$, $a_2^c=-0.317$, $a_1^b=1.14$ and $a_2^b=-0.20$ as in
\cite{Kiselev:2003mp,Brambilla:2004wf}. In Table~\ref{tab:Bc-nonlep1} we
give our results for the nonleptonic decays of the $B_c$ meson into two
ground state mesons and compare our results with the results of other model
calculations.

For the $b \to c$ induced decays our results are generally
close to the QCD sum rule results of \cite{KKL,exBc} and the constituent quark
model results of \cite{Chang:1992pt,narod,CdF,Faust,AbdEl-Hady:1999xh}. In
exception are the $(b \to c ; c \to (s,d))$ results of \cite{Chang:1992pt}
which are considerably smaller than our results, and smaller than the
results of the other model calculations. Summing up the exclusive
contributions one obtains a branching fraction of $8.8\%$. Considering the
fact that the $b \to c$ contribution to the total rate is expected to be
about 20\% \cite{Brambilla:2004wf} this leaves plenty of room for
nonresonant multibody decays

For the $c \to s$ induced decays our branching ratios are considerably
smaller than those predicted by QCD sum rules \cite{KKL,exBc} but are
generally close to the other constituent quark model results. When we sum
up our exclusive branching fractions we obtain a total branching ratio of
27.6\% which has to be compared with the 70\% expected for the $c \to s$
contribution to the total rate \cite{Brambilla:2004wf}. The sum rule model
of \cite{KKL,exBc} gives a summed branching fraction of 73.4\% for the
$c \to s$ contribution, i.e. the model of \cite{KKL,exBc} predicts that the
exclusive channels pretty well saturate the $c \to s$ part of the total rate.

Of interest are the ratios of branching ratios of the pairs of modes
$B_c \to VV$ and $B_c \to VP$, and $B_c \to VP$ and $B_c \to PP$ where one
expects from naive spin counting
that the rate ratios $VP/PP$ and $VV/VP$ are $\approx 3$. In many of the
pairs of decay modes naive spin counting can be seen to hold. However, for
some of the pairs one finds approximate equality or even an inversion of
the naive spin counting ratio. The deviation from naive spin counting can
be seen to be a common feature of all model results.

As was pointed out in \cite{CPBc} and further elaborated in
\cite{Ivanov:2002un,CPBcKis,CPLiu}
the decays $B^-_c\to D^-_s D^0(\overline{ D^0})$
are well suited for an extraction of the CKM angle $\gamma$ through
amplitude relations. These decays are better suited for the extraction
of $\gamma$ than the corresponding  decays of the $B_u$ and $B_d$ mesons
because  the unitarity triangles in the latter decays are rather squashed.
We list our updated results for these decays in Tables~VI and VII, where
Table~VI contains our results
for general values of the Wilson coefficients while Table~VII
list values of the branching ratios for specified values of the Wilson
coefficients. The branching ratios in Table~VII are quite small
${\cal O}(10^{-5})$
but these modes should still be accessible at high luminosity hadron colliders.

Finally, in Table~VIII, we compute the branching ratios
of the exclusive nonleptonic $B_c^-$ decays
into $p$--wave charmonium states, and into the orbital excitation of the
charmonium state $\psi(3836)$. We compare our results with the results of
other studies when they are available.

\begin{table}[t]
\caption{Leptonic decay constants $f_H$ (MeV) used in the least-squares
fit for our model parameters.}
\label{tab:leptonic}
\begin{center}
\def\arraystretch{1.5}
\begin{tabular}{|c|c|l|l|}
\hline
    & This work  & \hspace*{1cm} Other & \hspace*{1cm}Ref.  \\
\hline
$f_D$
 & 227 & $222.6\pm 16.7^{+2.8}_{-3.4}$  & CLEO \cite{Artuso:2005ym} \\
 &     & $201 \pm 3 \pm 17$            & MILC LAT \cite{Aubin:2005ar} \\
 &     & $235 \pm 8 \pm 14$            & LAT \cite{Chiu:2005ue} \\
 &     & $210 \pm 10^{+17}_{-16}$       & UKQCD LAT \cite{Lellouch:2000tw}\\
 &     & $211 \pm 14^{+2}_{-12}$        & LAT \cite{Becirevic:1998ua}\\
\hline
$f_{D^\ast}$
 & 249 & $245 \pm 20^{+3}_{-2}$ & LAT \cite{Becirevic:1998ua}\\
\hline
$f_{D_s}$
 & 255 & $266\pm 32 $                  & \cite{PDG} \\
 &     & $249 \pm 3 \pm 16$            & MILC LAT \cite{Aubin:2005ar} \\
 &     & $266 \pm 10\pm 18$            & LAT \cite{Chiu:2005ue}  \\
 &     & $290 \pm 20\pm 29\pm 29\pm 6$ & LAT \cite{Wingate:2003gm} \\
 &     & $236 \pm 8^{+17}_{-14}$        & UKQCD LAT \cite{Lellouch:2000tw}\\
 &     & $231 \pm 12^{+8}_{-1}$         & LAT \cite{Becirevic:1998ua}\\
$\displaystyle\frac{f_{D_s}}{f_D}$
 & 1.12 & $1.24 \pm 0.01\pm 0.07$         & MILC LAT \cite{Aubin:2005ar} \\
 &      & $1.13 \pm 0.03\pm 0.05$         & LAT \cite{Chiu:2005ue}  \\
 &      & $1.13 \pm 0.02^{+0.04}_{-0.02}$  & UKQCD LAT \cite{Lellouch:2000tw}\\
 &      & $1.10 \pm 0.02$                 & LAT \cite{Becirevic:1998ua}\\
\hline
$f_{D^\ast_s}$
 & 266 & $272 \pm 16^{+3}_{-20}$   &  LAT \cite{Aubin:2005ar} \\
\hline
$f_{\eta_c}$
 & 484 & $420\pm 52$ & \cite{Hwang:1997ie} \\
\hline
\end{tabular}
\begin{tabular}{|c|c|l|l|}
\hline
    & This work  & \hspace*{1cm} Other & \hspace*{1cm} Ref.  \\
\hline
$f_B$
 & 187  & $216\pm 9\pm 19\pm 4\pm 6$ & HPQCD LAT \cite{Gray:2005ad} \\
 &      & $177 \pm 17^{+22}_{-22}$    & UKQCD LAT \cite{Lellouch:2000tw}\\
 &      & $179 \pm 18^{+34}_{-9}$     & LAT \cite{Becirevic:1998ua}\\
 &&&\\
 &&&\\
\hline
$f_{B^\ast}$
 & 196  & $196 \pm 24^{+39}_{-2}$    & LAT \cite{Becirevic:1998ua}\\
\hline
$f_{B_s}$
 & 218 & $259\pm 32$                 & HPQCD LAT \cite{Gray:2005ad} \\
 &     & $260 \pm 7\pm 26\pm 8\pm 5$ & LAT \cite{Wingate:2003gm} \\
 &     & $204 \pm 12^{+24}_{-23}$     & UKQCD LAT \cite{Lellouch:2000tw}\\
 &     & $204 \pm 16^{+36}_{-0}$      & LAT \cite{Becirevic:1998ua}\\
 &&&\\
 &&&\\
$\displaystyle\frac{f_{B_s}}{f_B}$
 & 1.16 & $ 1.20\pm 0.03\pm 0.01 $        & HPQCD LAT \cite{Gray:2005ad} \\
 &      & $1.15 \pm 0.02^{+0.04}_{-0.02}$  & UKQCD LAT \cite{Lellouch:2000tw}\\
 &      & $1.14 \pm 0.03^{+0.01}_{-0.01}$  & LAT \cite{Becirevic:1998ua}\\
 &&&\\
\hline
$f_{B^\ast_s}$
 & 229  & $229 \pm 20^{+41}_{-16}$  & LAT \cite{Becirevic:1998ua}\\
\hline
$f_{B_c}$
 & 399 & $395 \pm 15$ & \cite{Kiselev:2003uk} \\
\hline
\end{tabular}
\end{center}
\end{table}


\begin{table}[t]
\caption{\label{tab:Bc-semlep1}
         Branching ratios (in $\%$) of
         exclusive semileptonic $B_c$ decays into ground state charmonium
         states, and into ground state charm and bottom meson states.
         For the lifetime of the $B_c$ we take $\tau(B_c) =0.45$ ps.}
\def\arraystretch{1.5}
\begin{center}
\begin{tabular}{|l|l|l|l|l|l|l|l|l|}
\hline
 Mode & This work  & \cite{KKL,exBc} & \cite{Chang:1992pt} &
 \cite{narod} & \cite{CdF} & \cite{Faust} & \cite{AbdEl-Hady:1999xh} &
\cite{Nobes:2000pm} \\
\hline
$B_c^- \to \eta_c e \nu$     & 0.81 & 0.75 & 0.97 & 0.59 &
 0.15 & 0.40 & 0.76 & 0.51 \\
$B_c^- \to \eta_c \tau \nu$  & 0.22 & 0.23 & -    & 0.20 &
-& -& -& -\\
\hline
$B_c^- \to J/\psi e \nu $    & 2.07 & 1.9  & 2.35 & 1.20 &
1.47 & 1.21 & 2.01 & 1.44 \\
$B_c^- \to J/\psi \tau \nu $ & 0.49 & 0.48 & -    & 0.34 &
-&- &- & -\\
\hline
 $B_c^- \to  \overline D^0 e \nu $       & 0.0035    & 0.004 & 0.006 & -  &
0.0003 & 0.001 & 0.003 & 0.0014\\
 $B_c^- \to  \overline D^0 \tau \nu $  & 0.0021    & 0.002 & -     & -  &
-& -& -& -\\
\hline
 $B_c^- \to  \overline D^{\ast\,0} e \nu  $ & 0.0038    & 0.018 & 0.018 & -  &
0.008 & 0.008 & 0.013 & 0.0023 \\
 $B_c^- \to  \overline D^{\ast\,0} \tau \nu$& 0.0022    & 0.008 & -     & -  &
 - & - & - & -\\
\hline\hline
 $B_c^- \to  \overline B^0_s e \nu  $  & 1.10    & 4.03  & 1.82  & 0.99 &
0.8 & 0.82 & 0.98 & 0.92 \\
 $B_c^- \to \overline B_s^{\ast\,0} e \nu$  & 2.37    & 5.06  & 3.01  & 2.30 &
2.3 & 1.71 & 3.45 & 1.41\\
\hline
 $B_c^- \to \overline B^0 e \nu  $    & 0.071    & 0.34  & 0.16  & -  &
0.06 & 0.04 & 0.078 & 0.048\\
 $B_c^- \to \overline B^{\ast\,0} e \nu  $  & 0.063    & 0.58  & 0.23  & -  &
0.19 & 0.12 & 0.24 & 0.051\\
\hline
\end{tabular}
\end{center}
\end{table}

\begin{table}[t]
\caption{\label{tab:Bc-semlep2}
         The branching ratios (in $\%$) of
         exclusive semileptonic $B_c^-$ decays into $p$--wave
         charmonium states, and into the $^3D_2$ orbital
         excitation of the charmonium state $\psi(3836)$.
         For the lifetime of the $B_c$ we take $\tau(B_c)= 0.45$ ps.}
\begin{center}
\begin{tabular}{|l|l|l|}
\hline\hline
 Mode & This work  & \cite{Chang:2001pm} \\
\hline
$B_c^-\to \chi_{c0}\, e\,\nu$     & 0.17   & 0.12  \\
$B_c^-\to\chi_{c0} \,\tau\,\nu$   & 0.013  & 0.017  \\
\hline
$B_c^-\to \chi_{c1} \, e\,\nu$    & 0.092   & 0.15 \\
$B_c^-\to \chi_{c1} \,\tau\,\nu$  & 0.0089  & 0.024\\
\hline
$B_c^-\to h_c\, e\,\nu $          & 0.27   & 0.17 \\
$B_c^-\to h_c\, \tau\,\nu$        & 0.017  & 0.024\\
\hline
$B_c^-\to \chi_{c2}\, e\,\nu $    & 0.17   & 0.19 \\
$B_c^-\to \chi_{c2}\, \tau\,\nu$  & 0.0082 & 0.029 \\
\hline
$B_c^-\to \psi(3836)\, e\,\nu $   & 0.0066 & - \\
$B_c^-\to \psi(3836)\, \tau\,\nu$ & 0.000099& -\\
\hline\hline
\end{tabular}
\end{center}
\end{table}


\begin{table}[th]
\caption{\label{tab:Bc-nonlepa1a2}
Exclusive nonleptonic decay widths
of the $B_c$ meson in units of $10^{-15}$ GeV
for general values of the Wilson coefficients $a_1$ and $a_2$. }
\begin{center}
\begin{tabular}{||l|l|}
\hline
 $B_c^- \to \eta_c \pi^-$       & $ 2.11\,a_1^2  $ \\
 $B_c^- \to \eta_c \rho^-$      & $ 5.10\,a_1^2 $  \\
 $B_c^- \to \eta_c K^- $        & $ 0.166\,a_1^2 $  \\
 $B_c^- \to \eta_c K^{\ast\,-}$  & $ 0.276\,a_1^2 $  \\
\hline
 $B_c^- \to J/\psi \pi^-$       & $ 1.93\, a_1^2 $  \\
 $B_c^- \to J/\psi \rho^-$      & $ 5.49\, a_1^2 $  \\
 $B_c^- \to J/\psi K^-$         & $ 0.150\, a_1^2 $  \\
 $B_c^- \to J/\psi K^{\ast\,-}$  & $ 0.313\, a_1^2 $ \\
\hline
 $B_c^- \to \chi_{c0}\, \pi^-$      &  $ 0.622\, a_1^2  $ \\
 $B_c^- \to \chi_{c0}\, \rho^-$     &  $ 1.47\, a_1^2 $  \\
 $B_c^- \to \chi_{c0}\, K^-$        &  $ 0.0472\, a_1^2 $  \\
 $B_c^- \to \chi_{c0}\, K^{\ast\,-}$ &  $ 0.0787\, a_1^2 $ \\
 \hline
 $B_c^- \to \chi_{c1}\, \pi^-$      &  $ 0.0768\, a_1^2 $  \\
 $B_c^- \to \chi_{c1}\, \rho^-$     &  $ 0.326\, a_1^2 $ \\
 $B_c^- \to \chi_{c1}\, K^-$          &  $ 0.00574\, a_1^2 $ \\
 $B_c^- \to \chi_{c1}\, K^{\ast\,-}$ &  $ 0.0201\, a_1^2 $  \\
\hline
 $B_c^- \to h_c\, \pi^-$            &  $ 1.24\, a_1^2 $  \\
 $B_c^- \to h_c\, \rho^-$           &  $ 2.78\, a_1^2 $ \\
 $B_c^- \to h_c\, K^-$              &  $ 0.0939\, a_1^2 $ \\
 $B_c^- \to h_c\, K^{\ast\,-}$       &  $ 0.146\, a_1^2 $ \\
\hline
 $B_c^- \to \chi_{c2}\, \pi^-$      &  $ 0.518\, a_1^2  $ \\
 $B_c^- \to \chi_{c2}\, \rho^-$     &  $ 1.33\, a_1^2  $ \\
 $B_c^- \to \chi_{c2}\, K^-$        &  $ 0.0384\, a_1^2  $ \\
 $B_c^- \to \chi_{c2}\, K^{\ast\,-}$ &  $ 0.0732\, a_1^2  $ \\
\hline
 $B_c^- \to \psi(3836)\, \pi^- $          &  $ 0.0193\, a_1^2 $  \\
 $B_c^- \to \psi(3836)\, \rho^-$          &  $ 0.0621\, a_1^2 $  \\
 $B_c^- \to \psi(3836)\, K^-   $          &  $ 0.00137\, a_1^2 $  \\
 $B_c^- \to \psi(3836)\, K^{\ast\,-}$      &  $ 0.00355\, a_1^2 $  \\
\hline\hline
 $B_c^- \to \eta_c D_s^{-}$            &  $  (2.73\, a_1+2.82\, a_2)^2  $ \\
 $B_c^- \to \eta_c D_s^{\ast\,-}$      &  $  (2.29\, a_1+1.51\,a_2)^2  $ \\
 $B_c^-  \rightarrow J/\psi D_s^-$     &  $  (2.19\, a_1+1.32\, a_2)^2  $ \\
 $B_c^-  \to J/\psi D_s^{\ast\,-}$     &  $  (3.69\, a_1+2.35\, a_2)^2  $ \\
\hline
 $B_c^- \to \eta_c D^-$           & $  (0.562\, a_1+0.582\, a_2)^2 $ \\
 $B_c^- \to \eta_c D^{\ast\,-}$   & $  (0.511\, a_1+0.310\, a_2)^2 $ \\
 $B_c^- \to J/\psi D^-$           & $  (0.462\, a_1+0.277\, a_2)^2 $ \\
 $B_c^- \to J/\psi D^{\ast\,-}$   & $  (0.785\, a_1+0.460\, a_2)^2 $ \\
\hline
\end{tabular}
\begin{tabular}{|l|l||}
\hline
 $B_c^- \to \overline B_s^0 \pi^-$          & $  39.7\, a_1^2 $ \\
 $B_c^- \to \overline B_s^0 \rho^-$         & $  23.6\, a_1^2 $ \\
 $B_c^- \to \overline B_s^{*0} \pi^-$       & $  21.8\, a_1^2 $ \\
 $B_c^- \to \overline B_s^{*0} \rho^-$      & $  115\, a_1^2 $ \\
 $B_c^- \to \overline B_s^0 K^-$            & $  2.93\, a_1^2 $ \\
 $B_c^- \to \overline B_s^{*0} K^-$         & $  1.34\, a_1^2 $ \\
 $B_c^- \to \overline B_s^0 K^{\ast\,-}$    & $  0.115\, a_1^2 $ \\
 $B_c^- \to \overline B_s^{*0}K^{\ast\,-}$  & $  5.12\, a_1^2 $ \\
\hline
 $B_c^- \to \overline B^0 \pi^-$            & $  2.04\, a_1^2 $ \\
 $B_c^- \to \overline B^0 \rho^-$           & $  2.05\, a_1^2 $ \\
 $B_c^- \to \overline B^{\ast\,0} \pi^-$    & $  0.578\, a_1^2 $ \\
 $B_c^- \to \overline B^{\ast\,0} \rho^-$   & $  3.07\, a_1^2 $ \\
 $B_c^- \to \overline B^0 K^-$              & $  0.153\, a_1^2 $ \\
 $B_c^- \to \overline B^0 K^{\ast\,-}$      & $  0.0490\, a_1^2 $ \\
 $B_c^- \to \overline B^{\ast\,0} K^-$      & $  0.0361\, a_1^2 $ \\
 $B_c^- \to \overline B^{\ast\,0} K^{\ast\,-}$   & $  0.133\, a_1^2 $ \\
\hline
 $B_c^- \to B^-  K^0$            & $  54.7\, a_2^2 $ \\
 $B_c^- \to B^-  K^{\ast\,0}$    & $  16.7\, a_2^2 $ \\
 $B_c^- \to B^{\ast\,-}  K^0$    & $  12.8\, a_2^2 $ \\
 $B_c^- \to B^{\ast\,-}  K^{\ast\,0}$ & $  46.3\, a_2^2 $ \\
 $B_c^- \to B^- \pi^0$                    & $  1.02\, a_2^2 $ \\
 $B_c^- \to B^- \rho^0$                   & $  1.03\, a_2^2 $ \\
 $B_c^- \to B^{\ast\,-} \pi^0$             & $  0.289\, a_2^2 $ \\
 $B_c^- \to B^{\ast\,-} \rho^0$            & $  1.54\, a_2^2 $ \\
\hline\hline
\multicolumn{2}{|c||}{$\displaystyle\frac{}{}$ }\\
\multicolumn{2}{|c||}{$\displaystyle\frac{}{}$ }\\
\multicolumn{2}{|c||}{$\displaystyle\frac{}{}$ }\\
\multicolumn{2}{|c||}{$\displaystyle\frac{}{}$ }\\
\multicolumn{2}{|c||}{$\displaystyle\frac{}{}$ }\\
\multicolumn{2}{|c||}{$\displaystyle\frac{}{}$ }\\
\multicolumn{2}{|c||}{$\displaystyle\frac{}{}$ }\\
\multicolumn{2}{|c||}{$\displaystyle\frac{}{}$ }\\
\multicolumn{2}{|c||}{$\displaystyle\frac{}{}$ }\\
\multicolumn{2}{|c||}{$\displaystyle\frac{}{}$ }\\
\hline
\end{tabular}
\end{center}
\end{table}

\begin{table}[t]
\caption{\label{tab:Bc-nonlep1}
         Branching ratios (in $\%$)
         of exclusive nonleptonic $B_c$ decays
         with the choice of Wilson coefficient:
         $a_1^c =1.20$ and $a_2^c=-0.317$ for c-decay,
         and $a_1^b =1.14$ and $a_2^b=-0.20$ for b-decay.
         For the lifetime of  the $B_c$ we take $\tau(B_c) = 0.45$ ps.}
\begin{center}
\begin{tabular}{|l|l|l|l|l|l|l|l|l|}
\hline
 Mode & This work  & \cite{KKL,exBc} & \cite{Chang:1992pt} &
 \cite{narod} & \cite{CdF} & \cite{Faust} & \cite{AbdEl-Hady:1999xh} \\
\hline
 $B_c^- \to \eta_c \pi^-$  & 0.19   & 0.20  & 0.18  & 0.13  &
0.025 & 0.083 & 0.14  \\
 $B_c^- \to \eta_c \rho^-$ & 0.45   & 0.42  & 0.49  & 0.30  &
0.067 & 0.20 & 0.33  \\
 $B_c^- \to \eta_c K^- $   & 0.015  & 0.013 & 0.014 & 0.013 &
0.002 & 0.006 & 0.011 \\
 $B_c^- \to \eta_c K^{\ast\,-}$ & 0.025  & 0.020 & 0.025 & 0.021 &
0.004 & 0.011 & 0.018  \\
\hline
 $B_c^- \to J/\psi \pi^-$  & 0.17   & 0.13  & 0.18  & 0.073 &
0.13 & 0.060 & 0.11 \\
 $B_c^- \to J/\psi \rho^-$ & 0.49   & 0.40  & 0.53  & 0.21  &
0.37 & 0.16 & 0.31  \\
 $B_c^- \to J/\psi K^-$    & 0.013  & 0.011 & 0.014 & 0.007 &
0.007 & 0.005 & 0.008  \\
 $B_c \to J/\psi K^{\ast\,-}$   & 0.028  & 0.022 & 0.029 & 0.016 &
0.020 & 0.010 & 0.018 \\
\hline
 $B_c^- \to \eta_c D_s^{-}$  & 0.44  & 0.28 & 0.054 & 0.35 &
0.50 & -& 0.26  \\
 $B_c^- \to \eta_c D_s^{\ast\,-}$ & 0.37  & 0.27 & 0.044 & 0.36 &
0.057 &  -& 0.24  \\
 $B_c^-  \to J/\psi D_s^-$    & 0.34  & 0.17 & 0.041 & 0.12 &
0.35 &  -& 0.15  \\
 $B_c^-  \to J/\psi D_s^{\ast\,-}$ & 0.97  & 0.67 & -     & 0.62 &
0.75 &  -& 0.55  \\
\hline
 $B_c^- \to \eta_c D^-$      & 0.019  & 0.015 & 0.0012 & 0.010  &
0.005 & -& 0.014 \\
 $B_c^- \to \eta_c D^{\ast\,-}$   & 0.019  & 0.010 & 0.0010 & 0.0055 &
0.003 & -& 0.013  \\
 $B_c^- \to J/\psi D^-$      & 0.015  & 0.009 & 0.0009 & 0.0044 &
0.013 & -& 0.009 \\
 $B_c^- \to J/\psi D^{\ast\,-}$   & 0.045  & 0.028 & -      & 0.010  &
0.023 & -& 0.028 \\
\hline\hline
 $B_c^- \to \overline B_s^0 \pi^-$     & 3.9  & 16.4 & 5.75 & 3.42 &
3.01 & 2.46 & 1.56 \\
 $B_c^- \to \overline B_s^0 \rho^-$    & 2.3  & 7.2  & 4.41 & 2.33 &
1.34 & 1.38 & 3.86 \\
 $B_c^- \to \overline B_s^{*0} \pi^-$  & 2.1  & 6.5  & 5.08 & 1.95 &
3.50 & 1.58 & 1.23 \\
 $B_c^- \to \overline B_s^{\ast\,0} \rho^-$ & 11  & 20.2 & 14.8 & 12.1 &
10.8 & 10.8 & 16.8 \\
 $B_c^- \to \overline B_s^0 K^-$       & 0.29  & 1.06 & 0.41 & -    &
0.21 & 0.21 & 0.17  \\
 $B_c^- \to \overline B_s^{\ast\,0} K^-$    & 0.13  & 0.37 & 0.29 & -    &
0.16 & 0.11 &  0.13  \\
 $B_c^- \to \overline B_s^0 K^{\ast\,-}$    & 0.011 & -    & -    & -    &
0.0043 & 0.0030 & 0.10 \\
 $B_c^- \to \overline B_s^{\ast\,0} K^{\ast\,-}$ & 0.50  & -    & -    & -    &
- & -& 1.14 \\
\hline
 $B_c^- \to \overline B^0 \pi^-$       & 0.20  & 1.06  & 0.32 & 0.15  &
0.19 & 0.10 & 0.10 \\
 $B_c^- \to \overline B^0 \rho^-$      & 0.20  & 0.96  & 0.59 & 0.19  &
0.15 & 0.13 & 0.28 \\
 $B_c^- \to \overline B^{*0} \pi^-$    & 0.057 & 0.95  & 0.29 & 0.077 &
0.24 & 0.026 & 0.076  \\
 $B_c^- \to \overline B^{\ast\,0} \rho^-$   & 0.30    & 2.57  & 1.17 & 0.67  &
0.85 & 0.67 & 0.89 \\
 $B_c^- \to \overline B^0 K^-$         & 0.015  & 0.07  & 0.025 & -    &
0.014 & 0.009 & 0.010 \\
 $B_c^- \to \overline B^0 K^{\ast\,-}$      & 0.0048 & 0.015 & 0.018 & -    &
0.003 & 0.004 & 0.012 \\
 $B_c^- \to \overline B^{\ast\,0} K^-$      & 0.0036& 0.055 & 0.019 & -    &
0.012 & 0.004 & 0.006  \\
 $B_c^- \to \overline B^{\ast\,0} K^{\ast\,-}$   & 0.013  & 0.058 & 0.037 & -    &
0.033 & 0.032 & 0.065 \\
\hline
 $B_c^- \to B^-  K^0$       & 0.38  & 1.98  & 0.66  & 0.17  &
- & 0.23 & 0.27   \\
 $B_c^- \to B^- K^{\ast\,0}$    & 0.11  & 0.43  & 0.47  & 0.095 &
- & 0.09 & 0.32 \\
 $B_c^- \to B^{\ast\,-} K^0$    &0.088  & 1.60  & 0.50  & 0.061 &
- & 0.10 & 0.16   \\
 $B_c^- \to B^{\ast\,-} K^{\ast\,0}$ & 0.32  & 1.67  & 0.97  & 0.57  &
- & 0.82 & 1.70 \\
 $B_c^- \to B^- \pi^0$                & 0.0070  & 0.037 & 0.011 & 0.007 &
-& 0.003 & 0.004 \\
 $B_c^- \to B^- \rho^0$               & 0.0071  & 0.034 & 0.020 & 0.009 &
-& 0.005 & 0.010  \\
 $B_c^- \to B^{\ast\,-} \pi^0$             & 0.0020  & 0.033 & 0.010 & 0.004 &
-& 0.001 & 0.003  \\
 $B_c^- \to B^{\ast\,-} \rho^0$            & 0.011  & 0.09  & 0.041 & 0.031 &
-& 0.023 & 0.031 \\
\hline\hline
\end{tabular}
\end{center}
\end{table}


\begin{table}[ht]
\caption{\label{tab:bcdda1a2}
Exclusive nonleptonic decay widths
of the $B_c$ meson into $DD$-mesons in units of $10^{-15}$ GeV.}
\begin{center}
\vspace{0.3cm}
\begin{tabular}{|c|c||c|c|}
\hline
$ B_c^-\to D^-\, D^0$                  & $ 1.19\,a_2^2 $ &
$ B_c^-\to D^-\, \overline D^0$      & $(0.0225\,a_1+0.0225\,a_2)^2$ \\
\hline
$ B_c^-\to D^-\, D^{\ast 0}$                & $ 1.38\,a_2^2 $ &
$ B_c^-\to D^-\, \overline D^{\ast\, 0}$  & $(0.0118\,a_1+0.0234\,a_2)^2$ \\
\hline
$ B_c^-\to D^{\ast\,-}\,  D^0$  & $ 0.323\,a_2^2 $  &
$ B_c^-\to D^{\ast\,-}\, \overline D^0$           & $(0.0243\,a_1+0.0117\,a_2)^2$ \\
\hline
$ B_c^-\to D^{\ast\,-}\, D^{\ast\, 0}$             & $ 0.770\,a_2^2 $ &
$ B_c^-\to D^{\ast\,-}\, \overline D^{\ast\, 0}$ & $(0.0181\,a_1+0.0181\,a_2)^2$ \\
\hline\hline
$ B_c^-\to D_s^-\, D^0$             & $ 0.0779\,a_2^2 $ &
$ B_c^-\to D_s^-\, \overline D^0$ & $(0.111\,a_1+0.109\,a_2)^2$ \\
\hline
$ B_c^-\to D_s^-\, D^{\ast\, 0}$             & $ 0.0881\,a_2^2 $ &
$ B_c^-\to D_s^-\, \overline D^{\ast\, 0}$ & $(0.0580\,a_1+0.114\,a_2)^2$ \\
\hline
$ B_c^-\to D_s^{\ast\,-}\,  D^0$            & $ 0.0236\,a_2^2 $  &
$ B_c^-\to D_s^{\ast\,-}\, \overline D^0$ & $(0.113\,a_1+0.0598\,a_2)^2$ \\
\hline
$ B_c^-\to D_s^{\ast\,-}\, D^{\ast\, 0}$  & $ 0.0574\,a_2^2 $ &
$ B_c^-\to D_s^{\ast\,-}\, \overline D^{\ast\, 0}$ & $(0.0871\,a_1+0.0932\,a_2)^2$ \\
\hline
\end{tabular}
\end{center}
\end{table}

\begin{table}[t]
\caption{\label{tab:Bc-DD}
         Branching ratios in units of $10^{-6}$ of the exclusive
         nonleptonic $B_c$ decays into $DD$-mesons.
         For the Wilson coefficients we choose
         $a_1^b=1.14$ and $a_2^b=-0.20$ relevant for the non-leptonic decays
         of the $\bar b$ quark. For the lifetime of the $B_c$ we take
         $\tau(B_c)= 0.45$ ps.}
\begin{center}
\begin{tabular}{|l|r|r|r|r|r|r|}
\hline
 Mode & This work & \cite{CPBcKis} & \cite{Chang:1992pt} & \cite{CPLiu} &
 \cite{CdF} & \cite{AbdEl-Hady:1999xh} \\
\hline
$B_c^- \to D^- D^{\hspace{1pt}\raisebox{-1pt}{$\scriptscriptstyle 0$}}$
 & 33 & 53 & 18 & 86 & 4.1 & 17 \\
$B_c^- \to D^- D^{\hspace{1pt}\raisebox{-1pt}{$\scriptscriptstyle *0$}}$
 & 38 & 75 & 19 & 75 &  3.6 & 21 \\
$B_c^- \to  D^{\scriptscriptstyle *-} D^{\hspace{1pt}\raisebox{-1pt}{$\scriptscriptstyle 0$}}$
 & 8.8    & 49 & 18 & 30 & 40 & 7.9\\
$B_c^- \to  D^{\scriptscriptstyle *-} D^{\hspace{1pt}\raisebox{-1pt}{$\scriptscriptstyle *0$}}$
 & 21  & 330 & 30 & 55 & 66 & 23 \\
$B_c^- \to D_s^- D^{\hspace{1pt}\raisebox{-1pt}{$\scriptscriptstyle 0$}}$
 & 2.1 & 4.8 & 0.93 & 4.6 & 0.27 & 1.13 \\
$B_c^- \to D_s^-D^{\hspace{1pt}\raisebox{-1pt}{$\scriptscriptstyle *0$}}$
 & 2.4 & 7.1 & 0.97 & 3.9 & 0.25 & 1.35 \\
$B_c^- \to  D_s^{\scriptscriptstyle *-} D^{\hspace{1pt}\raisebox{-1pt}{$\scriptscriptstyle 0$}}$
 & 0.65   & 4.5 & 0.91 & 1.8 & 2.38 & 0.55 \\
$B_c^- \to  D_s^{\scriptscriptstyle *-} D^{\hspace{1pt}\raisebox{-1pt}{$\scriptscriptstyle *0$}}$
 & 1.6 & 26 & 1.54 & 3.5 & 4.1 & 1.63\\
\hline
\end{tabular}
\begin{tabular}{|l|r|r|}
\hline
 Mode & This  work & \cite{CPBcKis}  \\
\hline
$B_c^- \to D^-\overline D^{\hspace{1pt}\raisebox{-1pt}{$\scriptscriptstyle 0$}}$
 & 0.31 & 0.32 \\
$B_c^- \to D^- \overline D^{\hspace{1pt}\raisebox{-1pt}{$\scriptscriptstyle*0$}}$
 & 0.052 & 0.28 \\
$B_c^- \to  D^{\scriptscriptstyle *-} \overline D^{\hspace{1pt}\raisebox{-1pt}{$\scriptscriptstyle 0$}}$
 & 0.44 & 0.40 \\
$B_c^- \to  D^{\scriptscriptstyle *-} \overline D^{\hspace{1pt}\raisebox{-1pt}{$\scriptscriptstyle*0$}}$
 & 0.20 & 1.59 \\
$B_c^- \to D_s^- \overline D^{\hspace{1pt}\raisebox{-1pt}{$\scriptscriptstyle 0$}}$
 & 7.4 & 6.6 \\
$B_c^- \to D_s^- \overline D^{\hspace{1pt}\raisebox{-1pt}{$\scriptscriptstyle*0$}}$
 & 1.3 & 6.3 \\
$B_c^- \to  D_s^{\scriptscriptstyle *-} \overline D^{\hspace{1pt}\raisebox{-1pt}{$\scriptscriptstyle 0$}}$
 & 9.3 & 8.5 \\
$B_c^- \to  D_s^{\scriptscriptstyle *-} \overline D^{\hspace{1pt}\raisebox{-1pt}{$\scriptscriptstyle *0$}}$
 & 4.5 & 40.4 \\
\hline
\end{tabular}
\end{center}
\end{table}

\begin{table}[t]
\caption{\label{tab:Bc-nonlep2}
         Branching ratios (in ($\%$))
         of the exclusive nonleptonic $B_c^-$ decays
         into $p$--wave
         charmonium states, and into the $^3D_2$ orbital
         excitation of the charmonium state $\psi(3836)$.
         The choice of Wilson coefficient is:
         $a_1^c =1.20$ and $a_2^c=-0.317$ for c-decays,
         and $a_1^b =1.14$ and $a_2^b=-0.20$ for b-decays.
         For the lifetime of  the $B_c$ we take $\tau(B_c) = 0.45$ ps.}
\begin{center}
\begin{tabular}{|l|r|r|r|r|}
\hline
 Mode & This work & \cite{Chang:2001pm} &\cite{Kiselev:2001zb} &
\cite{LopezCastro:2002ud}  \\
\hline
 $B_c^-\to \chi_{c0}\, \pi^-$    & 0.055   & 0.028     & 0.98   & -       \\
 $B_c^-\to \chi_{c1}\, \pi^-$    & 0.0068  & 0.007     & 0.0089 & -       \\
 $B_c^-\to h_c\, \pi^-$          & 0.11    & 0.05      & 1.60   & -       \\
 $B_c^-\to \chi_{c2}\, \pi^-$    & 0.046   & 0.025     & 0.79   & 0.0076  \\
 $B_c^-\to \psi(3836)\, \pi^-$   & 0.0017  & -         & 0.030  & -       \\
\hline
 $B_c^-\to \chi_{c0}\, K^-$      & 0.0042  & 0.00021   &-       &-        \\
 $B_c^-\to \chi_{c1}\, K^-$      & 0.00051 & 0.000052  &-       &-        \\
 $B_c^-\to h_c\, K^-$            & 0.0083  & 0.00038   &-       &-        \\
 $B_c^-\to \chi_{c2}\, K^-$      & 0.0034  & 0.00018   &-       & 0.00056 \\
 $B_c^-\to \psi(3836)\, K^-$     & 0.00012 & -         &-       &-        \\
\hline
\end{tabular}
\begin{tabular}{|l|r|r|r|r|}
\hline
 Mode & This work & \cite{Chang:2001pm} &\cite{Kiselev:2001zb} &
\cite{LopezCastro:2002ud} \\
\hline
 $B_c^-\to \chi_{c0}\, \rho^-$   & 0.13    & 0.072     & 3.29   &-        \\
 $B_c^-\to \chi_{c1}\, \rho^-$   & 0.029   & 0.029     & 0.46   &-        \\
 $B_c^-\to h_c\, \rho^-$         & 0.25    & 0.12      & 5.33   &-        \\
 $B_c^-\to \chi_{c2}\, \rho^-$   & 0.12    & 0.051     & 3.20   & 0.023   \\
 $B_c^-\to \psi(3836)\, \rho^-$  & 0.0055  & -         & 0.98   &-        \\
\hline
 $B_c^-\to \chi_{c0}\, K^{\ast-}$ & 0.0070  & 0.00039   &-       &-        \\
 $B_c^-\to \chi_{c1}\, K^{\ast-}$ & 0.0018  & 0.00018   &-       &-        \\
 $B_c^-\to h_c\, K^{\ast-}$       & 0.013   & 0.00068   &-       &-        \\
 $B_c^-\to \chi_{c2}\, K^{\ast-}$ & 0.0065  & 0.00031   &-       & 0.0013  \\
 $B_c^-\to \psi(3836)\, K^{\ast-}$& 0.00032 & -         &-       &-        \\
\hline
\end{tabular}
\end{center}
\end{table}

\clearpage

\section{Angular decay distributions for the decays of the $B_c$ meson
into $J/\psi$ modes}
\label{sec:angular}

The exclusive decays of the $B_c$-meson involving a $J/\psi$ meson
have an excellent experimental signature since the $J/\psi$
can be readily reconstructed from its leptonic decay channels
$J/\psi\to \mu^+\mu^-,e^+e^-$. In fact, decays of the $B_c$ into
$J/\psi$ modes have been among the discovery channels of the $B_c$.
In this section we write down
the complete angular decay distributions for the nonleptonic decays
$B_c^- \to J/\psi (\to l^+l^-)+ \rho^-(\to \pi^- \pi^0)$,
$B_c^- \to J/\psi (\to l^+l^-)+ \pi^-$ and
$B_c^- \to \eta_c + \rho^-(\to \pi^- \pi^0)$, and the semileptonic
decay $B_c^-\to J/\psi +l^-+\bar{\nu_l}$.

The experimental analysis of the angular decay distributions allows one to
learn more about the decay dynamics of the $B_c$ decays. The decay dynamics is
encapsuled in the helicity structure functions that multiply the angular
factors in the
decay distribution. Vice versa, the explicit form of the
decay distributions may be a useful input for writing event generators for
the decay process where one now has to make use of some theoretical input to
determine explicit values for the helicity structure functions.

We first discuss the nonleptonic cascade decay $B_c^- \to J/\psi
(\to l^+l^-)+ \rho^-(\to \pi^- \pi^0)$. The branching ratio of this mode
is predicted to be approximately three times the branching ratio of the decay
$B_c^- \to J/\psi (\to l^+l^-)+ \pi^-$ which has already been seen
\cite{Acosta:2005us}. It is not difficult to anticipate that the decay mode
$B_c^- \to J/\psi + \rho^-$ will be one of the next exclusive decay modes to
be seen in the very near future. The
decay mode $B_c^- \to J/\psi + \rho^-$ will afford an excellent
opportunity to take a more detailed look at the spin dynamics of the primary
weak decay process through an analysis of the joint angular decay
distributions of the second stage decays $J/\psi \to l^+l^-$ and
$\rho^-\to \pi^- \pi^0)$.

\begin{figure}
\begin{center}
\epsfig{figure=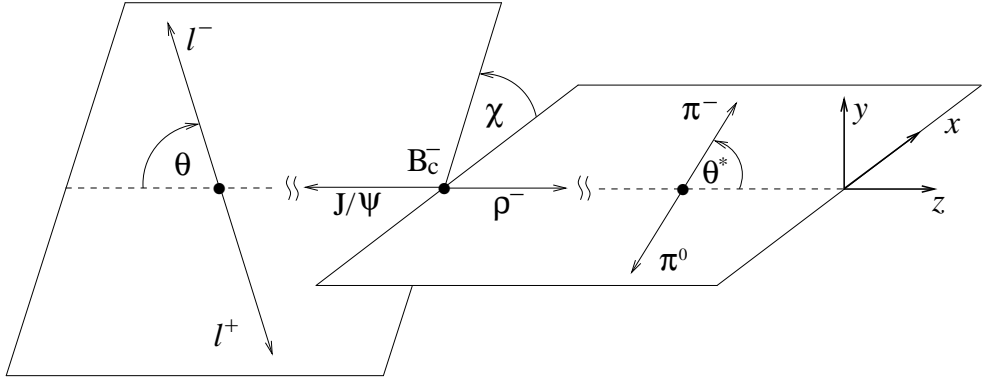,height=5cm}
\end{center}
\caption{Definition of the polar angles $\theta$ and
$\theta*$ and the azimuthal angle $\chi$ in the cascade decay
$B_c^- \to J/\psi (\to l^+l^-) + \rho^-(\to \pi^- \pi^0)$.}
\label{fig:angle}
\end{figure}

The angular decay distribution of the cascade decay
$B_c^- \to J/\psi (\to l^+l^-)+ \rho^-(\to \pi^- \pi^0)$ has been
discussed before in \cite{bkkz93,Kramer:1991xw,Dighe:1995pd,
Chiang:1999qn} including also lepton mass effects \cite{bkkz93}.
We rederive the angular decay distribution using the methods described in
\cite{ks90,Faessler:2002ut}. The angular decay distribution can be cast into
the form

\begin{eqnarray}
\label{angdist1}
\hspace{-0.5cm}
W(\theta, \chi, \theta^\ast) &\propto&
\sum_{\lambda=m,\lambda'=m' \atop \lambda_l,\lambda_{l'}}
|h_{\lambda_l\lambda_{l'}}|^2e^{i(m-m')(\pi-\chi)}\,
\times\,
d^1_{m,\lambda_l-\lambda_{l'}}(\theta)
d^1_{m',\lambda_l-\lambda_{l'}}(\theta)
H_{\lambda m}H^\dagger_{\lambda' m'}
d^1_{\lambda 0}(\theta^\ast)d^1_{\lambda' 0}(\theta^\ast)\,,
\end{eqnarray}
where the $d^1_{m m'}$ are Wigner's $d$--functions in the convention of Rose.
The summation in
(\ref{angdist1}) runs over $\lambda=m=0,\pm1$, $\lambda'=m'=0,\pm 1$ and
$\lambda_l,\lambda_{l'}=\pm 1/2$. The helicity amplitudes
$h_{\lambda_l\lambda_{l'}}$ describe the decay
$J/\psi(\lambda_l-\lambda_{l'}) \to l^+(\lambda_l)+l^-(\lambda_{l'})$
with lepton helicities $\lambda_l$ and $\lambda_{l'}$ where $l$ and $l'$
denote the positively and negatively charged leptons, respectively.
Similarly the
helicity amplitudes $H_{\lambda m}$ describe the decay
$B_c^- \to J/\psi(m)
+ \rho^-(\lambda)$ where the helicities of the $J/\psi$ and the $\rho^-$
are denoted by $m$ and $\lambda$, respectively.
Since $\lambda=m$ from angular momentum conservation
we shall drop one of the helicity labels in the
helicity amplitudes, i.e. we write $H_m$ for $H_{\lambda m}$
($m=0,\pm 1$). The angles $\theta, \chi$ and $\theta^\ast$
are defined in Fig.~\ref{fig:angle}.

We begin with by neglecting helicity flip effects in the decay
$J/\psi~\to~l^+l^-$, i.e. we take $\lambda_l=-\lambda_{l'}$
in Eq.~(\ref{angdist1}). Further we assume that the helicity amplitudes
are relatively real neglecting possible $T$--odd effects. One then obtains
\begin{eqnarray}
\label{angdist2}
\hspace{- 1.0cm}\frac{d \Gamma}{d\cos\theta\, d \chi\, d\cos\theta^\ast}
&=& \frac{1}{2\pi} \Big[\frac{3}{8}(1+\cos^2\theta)
\frac{3}{4}\sin^2\theta^\ast \Gamma_U
+\frac{3}{4}\sin^2\theta\frac{3}{2}\cos^2\theta^\ast \Gamma_L
\nonumber\\
&& + \frac{9}{32} \sin2\theta \cos\chi \sin2\theta^\ast \Gamma_I
- \frac{1}{2} \cdot \frac{3}{4}\sin^2\theta \cos2\chi
\frac{3}{4}\sin^2\theta^\ast \Gamma_T \Big]
\end{eqnarray}

Integrating (\ref{angdist2}) over $\chi$ and $\theta^\ast$ one obtains
\begin{equation}
\label{angdist3}
\frac{d \Gamma}{d\cos\theta}= \frac{3}{8}(1+\cos^2\theta)\Gamma_U
+\frac{3}{4} \sin^2\theta \Gamma_L :=\,\, \frac{3}{8}
(\Gamma_U+2\Gamma_L)(1+ \alpha_{L/T} \cos^2\theta)
\, ,
\end{equation}
where the asymmetry parameter
\begin{equation}
\label{alpha}
\alpha_{L/T} = \frac{\Gamma_U-2\Gamma_L}{\Gamma_U+2\Gamma_L}
\end{equation}
is a measure of the transverse/longitudinal composition of the produced
$J/\psi$.

Upon full angular integration one has $\Gamma=\Gamma_U+\Gamma_L$. Note that
we have taken the freedom to omit the branching ratio factors
${\rm Br}(J/\psi \to l^+l^-)$ and
${\rm Br}(\rho^- \to \pi^- \pi^0)$ on the right hand side of
Eq.~(\ref{angdist2}). The reason is that, upon angular integration, we want to
obtain the total rate $\Gamma(B_c^- \to J/\psi + \rho^-)$. The
partial rates $\Gamma_i(i=U,P,L,T,I)$ in Eq.~(\ref{angdist2}) are related to
bilinear products of the helicity amplitudes via
\begin{equation}
\Gamma_i = \frac{G_F^2}{16\pi}
|V_{cb} V_{ud} a_1 f_\rho m_\rho |^2
\frac{\bf |p_2|}{m_{B_c}^2} {\cal H}_i\,,
\end{equation}
where ${\bf |p_2|}$ is the magnitude of the three-momentum of the $\rho^-$
(or the $J/\psi$) in the rest frame of the $B_c$-meson. The helicity structure
functions ${\cal H}_i$ ($i=U,P,L,T,I$) are given by
\begin{eqnarray}
\label{hel_i}
{\cal H}_U &=& |H_+|^2 + |H_-|^2\,, \nonumber\\
{\cal H}_P &=& |H_+|^2 - |H_-|^2\,, \nonumber\\
{\cal H}_L &=& |H_0|^2\,,\\
{\cal H}_T &=& {\rm Re} H_+ H_-^\dagger\,, \nonumber\\
{\cal H}_I &=& \frac{1}{2} {\rm Re} (H_+H_0^\dagger +H_-H_0^\dagger ) \,. \nonumber
\end{eqnarray}
Using our constituent quark model results the partial rates $\Gamma_i$
take the following values
\begin{eqnarray}
\label{gam_i}
\Gamma_U&=&\, +0.826\cdot 10^{-15}\,{\rm GeV}, \nonumber\\
\Gamma_P&=&\, -0.644\cdot 10^{-15}\,{\rm GeV}, \nonumber\\
\Gamma_L&=&\, +6.30\cdot 10^{-15}\,{\rm GeV},  \\
\Gamma_T&=&\, +0.259\cdot 10^{-15}\,{\rm GeV}, \nonumber \\
\Gamma_I&=&\, +1.46\cdot 10^{-15}\,{\rm GeV}. \nonumber
\end{eqnarray}
Using the inverse lifetime $\tau(B_c)^{-1} = 1.463 \cdot 10^{-12}~$GeV and
the sum $\Gamma_U+\Gamma_L$ in Eq.~(\ref{gam_i}) one numerically reproduces
the branching ratio listed in Table \ref{tab:Bc-nonlep1}.

Even though ${\cal H}_P$ cannot be measured in the cascade decay
$B_c^- \to J/\psi (\to l^+l^-) + \rho^-(\to \pi^- \pi^0)$ we have included
the parity-odd helicity structure function
${\cal H}_P = |H_+|^2 - |H_-|^2$ in the results for illustrative reasons
in order to exemplify the hierarchy of helicity rates. The reason that
${\cal H}_P$ cannot be measured is that the
analyzing decays $J/\psi \to l^+l^-$ and $\rho^- \to \pi^- \pi^0$ are both
parity--conserving. From the numbers in Eq.~(\ref{gam_i}) one finds the
hierarchy $\Gamma_L/\Gamma : \Gamma_-/\Gamma : \Gamma_+/\Gamma =
88\% : 10.3\% : 1.3\%$
where $\Gamma_\pm= (\Gamma_U \pm \Gamma_P)/2$. The longitudinal rate
strongly dominates over the transverse rates. In terms of the asymmetry
parameter $\alpha_{L/T}$ defined in Eq.~(\ref{angdist3}) we find
$\alpha_{L/T}=-0.88$ as compared to
$\alpha_{L/T}=-0.85$ in \cite{Pakhomova:1999ky}. Among the transverse rates
the transverse--minus rate $\Gamma_-$ dominates over the transverse--plus
rate $\Gamma_+$ (for $\Gamma(B_c^+ \to J/\psi + \rho^+)$ one has
$\Gamma_+>\Gamma_-$). The predicted hierarchy of rates can be easily
understood in terms of simple spin arguments as were given some time ago in
\cite{Ali:1978kn, Korner:1979ci} and rediscovered in
\cite{Kagan:2004uw}. In the so--called $B \to VV$--decays the dominance of
the longitudinal mode has been experimentally confirmed in the decay
$B \to \rho \rho$ but not in the decay $B \to \phi K^\ast$ where enhanced
penguin effects may play an important role (see the discussion in
\cite{Kagan:2004uw,Ladisa:2004bp,Beneke:2005we}).

For the charge conjugate mode $B_c^+ \to J/\psi + \rho^+$ the angular
decay distributions in Eqs.~(\ref{angdist2}) and (\ref{angdist3}) will remain
unchanged since from $CP$ invariance one has
$\bar{H}_\lambda(B_c^+)=H_{-\lambda}(B_c^-)$ for real helicity amplitudes.
The partial helicity rates $\Gamma_{U,L,T,I}$ remain unchanged going from
the $B_c^-$ to the $B_c^+$ mode $(\bar{\Gamma}_{U,L,T,I}=\Gamma_{U,L,T,I})$
except for the
partial helicity rate $\Gamma_P$ which is not measurable in the decay.

For the sake of completeness we also list the corresponding angular decay
distribution when lepton mass and $T$--odd effects are included.
For nonvanishing lepton masses one now has to also include helicity flip
effects ($\lambda_l=\lambda_{l'}$)
in the decay $J/\psi~\to~l^+l^-$ which are nonvanishing for
nonvanishing lepton masses. We also drop the assumption that the helicity
amplitudes are relatively real thus including possible $T$--odd effects.

Although the helicity flip effects are expected to be quite small for the decay
$J/\psi~\to~\mu^+\mu^-$ we shall include them for completeness.
Lepton mass effects have to be taken into account e.g. in
the decay $J/\psi(2S) \to \tau^+ + \tau^-$ (not discussed in this paper)
since the $J/\psi(2S)$ has a
mass of 3686~MeV which lies above the $(\tau^+ + \tau^-)$--threshold
($2m_\tau=3.554~$GeV). When helicity flip effects are included one needs
to know the ratio of the squared flip and nonflip helicity amplitudes
of the decay $J/\psi~\to~\mu^+\mu^-$ which are given by
\begin{equation}
\label{flip-nonflip}
\frac{|h_{\frac{1}{2}\frac{1}{2}}|^2}{|h_{\frac{1}{2}-\frac{1}{2}}|^2}
= \frac{2m_l^2}{m_{J/\psi}^2} := 4\, \epsilon \,\,.
\end{equation}
The remaining two flip and nonflip amplitudes can be obtained from the
parity relations
$h_{-\frac{1}{2}-\frac{1}{2}}=h_{\frac{1}{2}\frac{1}{2}}$ and
$h_{-\frac{1}{2}\frac{1}{2}}=h_{\frac{1}{2}-\frac{1}{2}}$.

Using Eqs.~(\ref{angdist1}) and (\ref{flip-nonflip}) and putting in the
correct normalization one obtains (we now reinstitute the branching ratio
factors)
\begin{eqnarray}
\label{angdist4}
\hspace{- 1.0cm}\frac{d \Gamma}{d\cos\theta\, d \chi\, d\cos\theta^\ast}
&=&
{\rm Br}(J/\psi \to l^+l^-)
{\rm Br}(\rho^- \to \pi^- \pi^0)\frac{G_F^2}{16\pi}
|V_{cb} V_{ud} a_1 f_\rho m_\rho |^2
\frac{\bf |p_2|}{m_{B_c}^2} \frac{1}{1+4 \epsilon}
\\
&&\hspace{-3.0cm}
\times\frac{9}{64\pi}\Big[\,\left(|H_+|^2+|H_-|^2 \right)
    (1+\cos^2\theta)\sin^2\theta^\ast
  + 4\,|H_0|^2\sin^2\theta \cos^2\theta^\ast
\nonumber \\
&&\hspace{-2cm}
+\,\left(
{\rm Re}(H_{0}H_{+}^\dagger)+{\rm Re}(H_{0}H_{-}^\dagger)
\right)
\sin2\theta\sin2\theta^*\cos\chi
\nonumber \\
&&\hspace{-2cm}
- 2\, {\rm Re}(H_-H_+^\dagger) \sin^2\theta \sin^2\theta^* \cos2\chi
\nonumber \\
&&\hspace{-2cm}
+\,\left( {\rm Im}(H_{0}H_{+}^\dagger)
       -{\rm Im}(H_{0}H_{-}^\dagger)\right)
\sin2\theta\sin2\theta^\ast \sin\chi
\nonumber \\
&&\hspace{-2cm}
+ 2\, {\rm Im}(H_{-}H_{+}^\dagger) \sin^2\theta\sin^2\theta^\ast\sin2\chi
\nonumber \\
&&\hspace{-3.5cm}
+\frac{m_l^2}{2m_{J/\psi}^2}
\Big\{8\left(|H_{+}|^2 +|H_{-}|^2 \right)
\sin^2\theta\sin^2\theta^\ast
+ 32\,|H_{0}|^2 \sin^2\theta \cos^2\theta^\ast
\nonumber \\
&&\hspace{-2cm}
-8\, \left({\rm Re}(H_{0}H_{+}^\dagger)
          +{\rm Re}(H_{0}H_{-}^\dagger)\right)
\sin2\theta\sin2\theta^* \cos\chi
\nonumber \\
&&\hspace{-2cm}
+16\, {\rm Re}(H_{-}H_{+}^\dagger) \cos^2\theta \sin^2\theta^* \cos2\chi
\nonumber \\
&&\hspace{-2cm}
-8\,\left({\rm Im}(H_{0}H_{+}^\dagger)
         -{\rm Im}(H_{0}H_{-}^\dagger)\right)
\sin2\theta\sin2\theta^\ast \sin\chi
\nonumber \\
&&\hspace{-2cm}
-16\, {\rm Im}(H_{-}H_{+}^\dagger)
\sin^2\theta \sin^2\theta^\ast \sin2\chi
\,\,\Big\}\,\, \Big].
\nonumber
\end{eqnarray}
We have checked that the angular decay distribution in Eq.~(\ref{angdist4})
agrees with the corresponding angular decay distribution written down in
\cite{bkkz93}. In addition, we have checked the correctness of the signs
of the nonflip azimuthal correlations by going through a fully covariant
calculation. Note that the angular decay distribution is invariant under
$\theta \to \pi-\theta,\,\, \chi \to \chi+\pi$ and
$\theta^\ast \to \pi-\theta^\ast, \,\, \chi \to \chi+\pi$
showing that the polar and azimuthal angles in Fig.~\ref{fig:angle}
could have also been defined by changing the labels
$l^+ \leftrightarrow l^-$ and/or $\pi^- \leftrightarrow \pi^0$.

We have also included so-called T--odd contributions in the decay
distribution (\ref{angdist4}) which can have their origin in possible imaginary
parts of the helicity amplitudes. These could arise from strong interaction
phases generated from final state interaction effects or from weak phases
occurring in extensions of the Standard Model (see e.g.
\cite{Korner:1990yx,Korner:1992kj}). In the
Standard Model and in the factorization approximation these T--odd
contributions vanish, i.e. the angular decay distribution (\ref{angdist4})
would be reduced to that part given by the real contributions listed in
(\ref{angdist2}). It
would nevertheless be interesting to experimentally check on the possible
presence of T--odd contributions in the angular decay distribution
Eq.~(\ref{angdist4}).

The angular decay distribution for the charge conjugate mode
$B_c^+ \to J/\psi + \rho^+$ can be obtained from Eq.~(\ref{angdist4})
by the replacement
$H_i(B_c^-) \to \bar{H}_i(B_c^+)$ in Eq.~(\ref{angdist4}). The charge
conjugate helicity amplitudes $\bar{H}_i$ and the helicity amplitudes
$H_i$ are related by (see e.g. \cite{Kramer:1991xw,Korner:1992kj})
\begin{equation}
\displaystyle{
\begin{array}{lcl}
H_\pm = |H_\mp|e^{i(\delta_\pm +\phi_\pm)} & & H_0 = |H_0|e^{i(\delta_0 + \phi_0)}\\
\vspace{-0.3truecm}& & \\
\bar{H}_\pm = |H_\mp|e^{i(\delta_\pm -\phi_\pm)} & &  \bar{H}_0 = |H_0|e^{i(\delta_0 - \phi_0)}
\end{array}}
\end{equation}
where the $\delta_i$ and $\phi_i$ denote the strong and weak phases of the
helicity amplitudes, respectively. A discussion of $CP$ violating
observables in this process can be found in \cite{Kramer:1991xw,Dighe:1995pd,
Chiang:1999qn}.

Next we turn to the angular decay distribution for the decay
$B_c^- \to J/\psi (\to l^+l^-)+ \pi^- $. It can be obtained from
Eq.~(\ref{angdist4}) by setting the transverse
helicity amplitudes to zero and replacing the longitudinal helicity amplitude
$H_{0}$ by the corresponding scalar (or time-component) helicity amplitude
$H_t$ of the decay $B_c \to J/\psi + \pi^- $. After $\cos\theta$-- and
$\chi$--integration one obtains
\begin{eqnarray}
\label{angdist5}
\hspace{- 1.0cm}\frac{d \Gamma}{d\cos\theta} &=&
{\rm Br}(J/\psi \to l^+l^-)\frac{G_F^2}{16\pi}
|V_{cb} V_{ud} a_1 f_\pi m_\pi |^2
\frac{\bf |p_2|}{m_{B_c}^2} \frac{1}{1+4 \epsilon}
|H_{t}|^2\,\,\frac{3}{4}(\sin^2\theta + 8 \epsilon \cos^2\theta)\,.
\end{eqnarray}
In a similar way one obtains the angular decay distribution for the decay
$B_c^- \to \eta_c + \rho^-(\to \pi^- \pi^0)$ where one finds
\begin{eqnarray}
\label{angdist6}
\hspace{- 1.0cm}\frac{d \Gamma}{d\cos\theta^\ast} &=&
{\rm Br}(\rho^- \to \pi^- \pi^0)\frac{G_F^2}{16\pi}
|V_{cb} V_{ud} a_1 f_\rho m_\rho |^2
\frac{\bf |p_2|}{m_{B_c}^2}
|H_{0}|^2\,\,\frac{3}{2} \cos^2\theta^\ast\,\, ,
\end{eqnarray}
and where now $H_{0}$ is the helicity amplitude of the decay
$B_c^- \to \eta_c + \rho^-$.

\begin{figure}
\begin{center}
\epsfig{figure=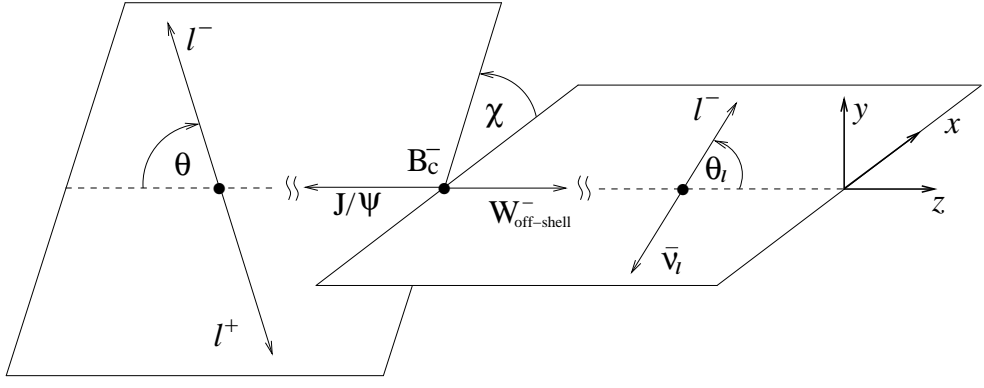,height=5cm}
\end{center}
\caption{Definition of the polar angles $\theta$ and
$\theta_l$ and the azimuthal angle $\chi$ in the cascade decay
$B_c^- \to J/\psi (\to l^+l^-) + W^-_{\rm{off-shell}} (\to l^- + \bar{\nu}_l)$.}
\label{fig:angle2}
\end{figure}

Finally we analyze the angular decay distribution in the semileptonic decay
$B_c^- \to J/\psi (\to l^+l^-) + l^- + \bar{\nu}_l$. We shall now
neglect lepton mass effects altogether and assume that the helicity
amplitudes are relatively real thus neglecting $T$--odd effects in the decay.
Using again the methods described in \cite{ks90,Faessler:2002ut} the angular
decay distribution can be cast into the form
\begin{eqnarray}
\label{angdist7}
\hspace{-0.5cm}
W(\theta, \chi, \theta_l) &\propto&
\sum_{\lambda=m,\lambda'=m' \atop \lambda_l=-\lambda_{\nu_l}}
e^{i(m-m')(\pi-\chi)}\,
\times\,
d^1_{m,\lambda_l-\lambda_{l'}}(\theta)
d^1_{m',\lambda_l-\lambda_{l'}}(\theta)
H_{\lambda m}H^\dagger_{\lambda' m'}
d^1_{\lambda, \mp1}(\theta_l)d^1_{\lambda', \mp1}(\theta_l)\,,
\end{eqnarray}
where the angles of the decay process are defined in Fig.~\ref{fig:angle2}.
Putting in the correct normalization one obtains
\begin{eqnarray}
\label{angdist8}
\hspace{- 1.0cm}\frac{d \Gamma}{dq^2d\cos\theta\, d \chi\, d\cos\theta_l}
= \frac{1}{2\pi}\bigg[&&\frac{3}{8}(1+\cos^2\theta)
\frac{3}{8}(1+\cos^2\theta_l) \frac{d \Gamma_U}{d q^2}
\\
&+&\frac{3}{4}\sin^2\theta\frac{3}{4}\sin^2\theta_l \frac{d \Gamma_L}{d q^2}
\nonumber \\
&\pm&\frac{3}{8}(1+\cos^2\theta)\frac{3}{4}\cos\theta_l
\frac{d \Gamma_P}{d q^2}
\nonumber \\
&+&\frac{3}{4}\sin^2\theta\frac{3}{4}\sin^2\theta_l\cos2\chi
\,\,2\frac{d \Gamma_T}{d q^2}
\nonumber \\
&+&\frac{3}{4}\sin2\theta\frac{3}{8}\sin2\theta_l\cos\chi
\frac{d \Gamma_I}{d q^2}
\nonumber \\
&\pm& \frac{3}{4}\sin2\theta\frac{3}{4}\sin\theta_l\cos\chi
\frac{d \Gamma_A}{d q^2} \bigg] \nonumber\,.
\end{eqnarray}

The contributions proportional to $\Gamma_P$ and $\Gamma_A$ in
Eq.~(\ref{angdist8}) change signs when going from the
$(l^-,\bar{\nu}_l)$ to the $(l^+,\nu_l)$ case, i.e. when going from the
decay $B_c^- \to J/\psi + l^- +\bar{\nu}_l$ to the decay
$B_c^+ \to J/\psi + l^+ +\nu_l$. For the terms proportional to $\Gamma_P$
and $\Gamma_A$ the upper
and lower signs holds for the $(l^-,\bar{\nu}_l)$ and
$(l^+,\nu_l)$ cases, respectively. However, since $\Gamma_P$ and $\Gamma_A$
also change signs when going from $B_c^- \to J/\psi + l^- + \bar{\nu}_l$
to $B_c^+ \to J/\psi + l^+ + \nu_l$ the form of the effective decay
distribution will be the same in both cases. Similar to the nonleptonic
decay $B_c^- \to J/\psi + \rho^-$ discussed earlier in this section
the angular decay distribution (\ref{angdist8}) is invariant under
$\theta \to \pi-\theta,\,\, \chi \to \chi+\pi$ showing that the polar
angle $\theta$ and the azimuthal angle $\chi$ in Fig.~\ref{fig:angle2} could have also been
defined by changing the labels $l^+ \leftrightarrow l^-$ in the decay
$J/\psi \to l^+l^-$.

Upon angular integration one has $d\Gamma/dq^2=d\Gamma_U/dq^2+d\Gamma_L/dq^2$.
Note that we have again taken the freedom to omit the branching ratio factor
${\rm Br}(J/\psi \to l^+l^-)$. The reason is again, when integrating
Eq.~(\ref{angdist8}) over $q^2$ and doing the angular integrations, we want to
obtain the total rate $\Gamma(B_c^- \to J/\psi + l + \nu_l)$.
After $\chi$ and $\cos\theta$ integration one recovers the corresponding
single angle decay distribution written down in \cite{Ivanov:2005fd}.
The differential partial helicity rates $ d \Gamma_i / d q^2$
$(i=U,L,P,T,I,A)$ are defined by
\begin{equation}
\label{partialrates}
\frac{d\Gamma_i}{dq^2} = \frac{G^2_F}{(2\pi)^3} \,
|V_{bc}|^2\,\frac{q^2\,|{\bf p_2}|}{12\,m_1^2\,}\,{\cal H}_i \, ,
\end{equation}
where
\begin{eqnarray}
{\cal H}_U &=& |H_+|^2 + |H_-|^2\,, \nonumber\\
{\cal H}_P &=& |H_+|^2 - |H_-|^2\,, \nonumber \\
{\cal H}_L &=& |H_0|^2\,,           \nonumber \\
{\cal H}_T &=& {\rm Re} H_+ H_-^\dagger\,,  \\
{\cal H}_I &=& \frac{1}{2} Re (H_+H_0^\dagger +H_-H_0^\dagger )\,,\nonumber \\
{\cal H}_A &=& \frac{1}{2} Re (H_+H_0^\dagger -H_-H_0^\dagger )\nonumber \,.
\end{eqnarray}
Numerically, we obtain the following values for the integrated partial
rates $\Gamma_i$
\begin{eqnarray}
\Gamma_U&=& +14.49 \cdot 10^{-15}\,{\rm GeV},\nonumber \\
\Gamma_P&=& -8.182 \cdot 10^{-15}\,{\rm GeV},\nonumber \\
\Gamma_L&=& +15.80 \cdot 10^{-15}\,{\rm GeV},\nonumber \\
\Gamma_T&=&  +5.850 \cdot 10^{-15}\,{\rm GeV},\\
\Gamma_I&=&  +9.494 \cdot 10^{-15}\,{\rm GeV},\nonumber \\
\Gamma_A&=& -3.208 \cdot 10^{-15}\,{\rm GeV}.\nonumber
\end{eqnarray}
The longitudinal rate $\Gamma_L$ and the (unpolarized) transverse rate
$\Gamma_U$ are of approximately equal size where the longitudinal rate
dominates at small $q^2$ (e.g. at $q^2=m_{\rho}^2$ as discussed earlier for
the decay $B_c^- \rightarrow J/\psi + \rho^-$)  and the transverse rate
dominates at large $q^2$. This implies that one no longer has a pronounced
longitudinal dominance in $B_c \to VV$ decays when the form factors are
probed at higher momentum transfers as e.g. in the decay
$B_c^- \rightarrow J/\psi +D^{\ast\,-} $. Reexpressing $\Gamma_U$ and
$\Gamma_P$
in terms of the transverse--minus and transverse--plus rates one finds
$\Gamma_-= 11.34\,\times 10^{-15}\,{\rm GeV}$ and
$\Gamma_+ =3.155\,\times 10^{-15}\,{\rm GeV}$. The dominance of $\Gamma_-$ over
$\Gamma_+$ reflects the basic left--chiral current structure of the
$b \to c$ current transition. The interference contributions $\Gamma_T$,
$\Gamma_I$ and $\Gamma_A$ are large enough to provide significant
azimuthal correlations in the semileptonic decay process.

\section{Summary and conclusions}
\label{sec:summary}
We have performed a comprehensive analysis of the exclusive semileptonic
and nonleptonic decays of the $B_c$--meson. The predicted branching ratios
range from very small numbers of ${\cal O}(10^{-6})$ up to the largest
branching fraction of 11$\%$ for the nonleptonic decay
$B_c^- \to \overline B_s^{\ast\,0} \rho^-$. We have compared our results with the
results of other studies. In general the results of the various model
calculations are of the same order of magnitude while they can differ
by factors of ten for specific decay modes. As a curious by--note we
mention that a first attempt at estimating exclusive nonleptonic $B_c$
decays can be found in \cite{Ali:1978kn}. Using the present value of
$V_{bc}$ (which was not known in 1978) and the present $\tau(B_c)= 0.45$ ps
the authors of \cite{Ali:1978kn}
calculated branching ratios of 0.29\% and 0.69\% for
$B_c \to J/\psi + \rho$ and $B_c \to J/\psi + \pi$, respectively,
with a $L:T_-:T_+$ ratio of $88\%:10.4\%:2\%$ for $B_c \to J/\psi + \rho$.
The branching ratios are $\approx 50\%$ above the branching ratios of the
present calculation, whereas the helicity rate composition of the decay
$B_c \to J/\psi + \rho$ is very close to that of the present model given in
Sec.~V. At any rate, we are looking forward to
a detailed experimental study of the many exclusive decay modes of the
$B_c$ meson described in this paper.

We have taken a more detailed look at the spin dynamics of the decay modes
$B_c^- \rightarrow J/\psi (\to l^+l^-) + \rho^-(\to \pi^- \pi^0)$
and $B_c^- \to J/\psi (\to l^+l^-) + W^-_{\rm{off-shell}}
(\to l^- + \bar{\nu}_l)$ involving a $J/\psi$ in the final state for which
we have presented explicit formulas for their joint angular decay
distributions. We have discussed the changes in the decay distributions
for the corresponding $B_c^+$ decay modes. It should be possible to test
the joint angular decay distributions and extract values for the helicity
structure functions with data samples of the ${\cal O}(100)$.

With our model assumptions the total exclusive rates calculated in this paper for the
$B_c^-$ decays are identical to the corresponding rates for the $B_c^+$ rates.
As concerns the partial transverse helicity rates in the $B_c \to VV$ modes
one has to change $T_- \leftrightarrow T_+$ when going from $B_c^- \to VV$ to
$B_c^+ \to VV$ as discussed in Sec.~\ref{sec:angular}.

\section*{Acknowledgments}
\noindent
M.A.I. appreciates the partial support by the DFG
(Germany) under the grant 436 RUS 17/26/04, the Heisenberg-Landau Program
and the Russian Fund of Basic Research (Grant No.04-02-17370).


\appendix
\section*{Appendix: Widths formulas for weak $B_c$-decays}

The leptonic decay widths are given by
\begin{eqnarray*}
\Gamma(P\to l \bar\nu) &=& \frac{G_F^2}{8\,\pi} |V_{q_1q_2}|^2
\,f_P^2\,m_P\,m_l^2\left[1-\frac{m_l^2}{m_P^2}\right]^2,
\\
&&\\
\Gamma(V\to  l \bar\nu) &=& \frac{G_F^2}{4\,\pi} |V_{q_1q_2}|^2
\,f_V^2\,m_V^3\left[1-\frac{m_l^2}{m_V^2}\right]^2
\left[1+\frac{m_l^2}{2\,m_V^2}\right].
\end{eqnarray*}
\noindent
For the semileptonic $B_c$-decay widths one finds

\begin{eqnarray*}
\Gamma(B_c^-\to M_{\bar cc}\, l \bar\nu) & = & \frac{G_F^2}{(2\,\pi)^3}
|V_{cb}|^2
\int\limits_{m_l^2}^{q^2_-} dq^2
\frac{(q^2-m_l^2)^2\,|{\mathbf p_2}|}{12\,m_1^2\,q^2}
\\
&&
\times\left\{
\left(1+\frac{m_l^2}{2\,q^2}\right)
\sum\limits_{i=\pm,0} \left(H_i^{B_c\to M_{\bar cc}}(q^2)\right)^2
+\frac{3\,m_l^2}{2\,q^2}\left(H_t^{B_c\to M_{\bar cc}}(q^2)\right)^2
\right\}\,,
\\
%
\Gamma(B_c^-\to \overline D^0\, l \bar\nu) & = & \frac{G_F^2}{(2\,\pi)^3}
|V_{ub}|^2
\int\limits_{m_l^2}^{q^2_-} dq^2
\frac{(q^2-m_l^2)^2\,|{\mathbf p_2}|}{12\,m_1^2\,q^2}
\\
&&
\times\left\{
\left(1+\frac{m_l^2}{2\,q^2}\right)
\sum\limits_{i=\pm,0} \left(H_i^{B_c\to \overline D^0}(q^2)\right)^2
+\frac{3\,m_l^2}{2\,q^2}\left(H_t^{B_c\to \overline D^0}(q^2)\right)^2
\right\}\,,
\\
%
\Gamma(B_c^-\to \overline B_q^0\, l \bar\nu) & = & \frac{G_F^2}{(2\,\pi)^3}
|V_{cd}|^2
\int\limits_{m_l^2}^{q^2_-} dq^2
\frac{(q^2-m_l^2)^2\,|{\mathbf p_2}|}{12\,m_1^2\,q^2}
\\
&&
\times\left\{
\left(1+\frac{m_l^2}{2\,q^2}\right)
\sum\limits_{i=\pm,0} \left(H_i^{B_c\to \overline B_q^0}(q^2)\right)^2
+\frac{3\,m_l^2}{2\,q^2}\left(H_t^{B_c\to \overline B_q^0}(q^2)\right)^2
\right\},
\\
&&
(q=s,d),
\end{eqnarray*}
where $q^2_{\pm}=(m_1\pm m_2)^2$, $m_1\equiv m_{B_c}$, and
$m_2\equiv m_f$. Note that $\overline D^0$ and $\overline B^0_q$ denote both
the pseudoscalar and vector cases.

\subsection{Nonleptonic $B_c$-Decay Widths}

Finally, the nonleptonic $B_c$-decay widths are given by the expressions in the following two sections.

\subsubsection{Transitions due to b-decays}
\begin{eqnarray*}
%
\Gamma(B^-_c\to P^- M_{\bar cc}) & = &
\frac{G^2_F}{16\,\pi}\frac{ |{\mathbf p_2}|}{m^2_1}
\left|V_{cb}V_{uq}^\dagger a_1 f_P m_P\right|^2
\left(H_t^{B_c\to M_{\bar cc}}(m^2_{P})\right)^2,
\\
&&
(P^-=\pi^-,K^{+},\,\,{\rm and}\,\, q=d,s,\,\,{\rm respectively}),
\\
&&\\
\Gamma(B^-_c\to V^- M_{\bar cc}) & = &
\frac{G^2_F}{16\,\pi} \frac{ |{\mathbf p_2}|}{m^2_1}
|V_{cb}V_{uq}^\dagger a_1 f_V m_V|^2
\sum\limits_{i=0,\pm}\left(H_i^{B_c\to M_{\bar cc}}(m^2_{V})\right)^2,
\\
&&
(V^-=\rho^-,K^{\ast\,+},\,\,{\rm and}\,\, q=d,s,\,\,{\rm respectively}),
\\
&&\\
\Gamma(B^-_c\to D_q^- \overline D^0) & = &
\frac{G^2_F}{16\,\pi} \frac{ |{\mathbf p_2}|}{m^2_1}
|V_{ub}V_{cq}^\dagger |^2
\left\{
a_1 f_{D^-_q}m_{D^-_q} H_t^{B_c\to \overline D^0}(m^2_{D^-_q})
+
a_2 f_{\overline D^0}m_{\overline D^0} H_t^{B_c\to D^-_q}(m^2_{\overline D^0})
\right\}^2
\\
&&\\
\Gamma(B^-_c\to D_q^{\ast\,-} \overline D^0) & = &
\frac{G^2_F}{16\,\pi} \frac{ |{\mathbf p_2}|}{m^2_1}
|V_{ub}V_{cq}^\dagger |^2
\left\{
a_1 f_{D^{\ast\,-}_q}m_{D^{\ast\,-}_q}
  H_0^{B_c\to \overline D^0}(m^2_{D^{\ast\,-}_q}) +
a_2 f_{\overline D^0}m_{\overline D^0} H_t^{B_c\to D^{\ast\,-}_q}(m^2_{\overline D^0})
\right\}^2
\\
&&\\
\Gamma(B^-_c\to D_q^- \overline D^{\ast\,0}) & = &
\frac{G^2_F}{16\,\pi} \frac{ |{\mathbf p_2}|}{m^2_1}
|V_{ub}V_{cq}^\dagger |^2
\left\{
a_1 f_{D^-_q} m_{D^-_q}
  H_t^{B_c\to \overline D^{\ast\,0}}(m^2_{D^-_q})
+
a_2 f_{\overline D^{\ast\,0}}m_{\overline D^{\ast\,0}} H_0^{B_c\to D^-_q}(m^2_{\overline D^{\ast\,0}})
\right\}^2
\\
&&\\
\Gamma(B^-_c\to D_q^{\ast\,-} \overline D^{\ast\,0}) & = &
\frac{G^2_F}{16\,\pi} \frac{ |{\mathbf p_2}|}{m^2_1}
|V_{ub}V_{cq}^\dagger |^2
\\
&&
\times\sum\limits_{i=0,\pm}
\left\{a_1 f_{D^{\ast\,-}_q} m_{D^{\ast\,-}_q}
H_i^{B_c\to \overline D^{\ast\,0}}(m^2_{D^{\ast\,-}_q})
     +a_2 f_{\overline D^{\ast\,0}}m_{\overline D^{\ast\,0}}
H_i^{B_c\to D^{\ast\,-}_q}(m^2_{\overline D^{\ast\,0}})
\right\}^2
\\
&&\\
&&\\
\Gamma(B^-_c\to D_q^{-} D^0) & = &
\frac{G^2_F}{16\,\pi} \frac{ |{\mathbf p_2}|}{m^2_1}
|V_{cb}V_{uq}^\dagger a_2 f_{ D^0} m_{ D^0} |^2
\left(H_t^{B_c\to D^{-}_q}(m^2_{ D^0})\right)^2
\\
&&\\
\Gamma(B^-_c\to D_q^{\ast\,-} D^0) & = &
\frac{G^2_F}{16\,\pi} \frac{ |{\mathbf p_2}|}{m^2_1}
|V_{cb}V_{uq}^\dagger a_2 f_{ D^0} m_{D^0} |^2
\left(H_t^{B_c\to D^{\ast\,-}_q}(m^2_{D^0})\right)^2
\\
&&\\
\Gamma(B^-_c\to D_q^- D^{\ast\,0}) & = &
\frac{G^2_F}{16\,\pi} \frac{ |{\mathbf p_2}|}{m^2_1}
|V_{cb}V_{uq}^\dagger a_2 f_{D^{\ast\,0}} m_{ D^{\ast\,0}} |^2
\left(H_0^{B_c\to D^{-}_q}(m^2_{D^{\ast\,0}})\right)^2
\\
&&\\
\Gamma(B^-_c\to D_q^{\ast\,-} D^{\ast\,0}) & = &
\frac{G^2_F}{16\,\pi} \frac{ |{\mathbf p_2}|}{m^2_1}
|V_{cb}V_{uq}^\dagger a_2 f_{D^{\ast\,0}} m_{D^{\ast\,0}} |^2
\sum\limits_{i=0,\pm}
\left(H_i^{B_c\to D^{\ast\,-}_q}(m^2_{D^{\ast\,0}})
\right)^2
\\
%
%
&&\\
\Gamma(B^-_c\to D_q^- \eta_c) & = &
\frac{G^2_F}{16\,\pi} \frac{ |{\mathbf p_2}|}{m^2_1}
|V_{cb}V_{cq}^\dagger |^2
\left\{
a_1 f_{D^-_q} m_{D^-_q}  H_t^{B_c\to \eta_c}(m^2_{D^-_q})
+
a_2 f_{\eta_c}m_{\eta_c} H_t^{B_c\to D^-_q}(m^2_{\eta_c})
\right\}^2
\\
&&\\
\Gamma(B^-_c\to D_q^{\ast\,-} \eta_c) & = &
\frac{G^2_F}{16\,\pi} \frac{ |{\mathbf p_2}|}{m^2_1}
|V_{cb}V_{cq}^\dagger |^2
\left\{
a_1 f_{D^{\ast\,-}_q} m_{D^{\ast\,-}_q}
  H_0^{B_c\to \eta_c}(m^2_{D^{\ast\,-}_q})
+
a_2 f_{\eta_c}m_{\eta_c} H_t^{B_c\to D^{\ast\,-}_q}(m^2_{\eta_c})
\right\}^2
\\
&&\\
\Gamma(B^-_c\to D_q^{-} J/\psi) & = &
\frac{G^2_F}{16\,\pi} \frac{ |{\mathbf p_2}|}{m^2_1}
|V_{cb}V_{cq}^\dagger |^2
\left\{
a_1 f_{D^-_q} m_{D^-_q}
  H_t^{B_c\to J/\psi }(m^2_{D^-_q})
+
a_2 f_{J/\psi}m_{J/\psi} H_t^{B_c\to D^{-}_q}(m^2_{J/\psi})
\right\}^2
\\
&&\\
\Gamma(B^-_c\to D_q^{\ast\,-} J/\psi)) & = &
\frac{G^2_F}{16\,\pi} \frac{ |{\mathbf p_2}|}{m^2_1}
|V_{cb}V_{cq}^\dagger |^2
\\
&&
\times\sum\limits_{i=0,\pm}
\left\{
 a_1 f_{D^{\ast\,-}} m_{D_q^{\ast\,-}}
 H_i^{B_c\to J/\psi}(m^2_{D_q^{\ast\,-}})
+
 a_2 f_{J/\psi} m_{J/\psi}
 H_i^{B_c\to D_q^{\ast\,-}}(m^2_{J/\psi})
\right\}^2
\end{eqnarray*}

\subsubsection{Transitions due to c-decays}
\begin{eqnarray*}
\Gamma(B^-_c\to \overline B^0 P^-) & = &
\frac{G^2_F}{16\,\pi} \frac{ |{\mathbf p_2}|}{m^2_1}
|V_{cd}V_{uq}^\dagger a_1 f_{P} m_{P} |^2
\left(H_t^{B_c\to \overline B^0}(m^2_{P})\right)^2
\\
\Gamma(B^-_c\to \overline B^0 V^-) & = &
\frac{G^2_F}{16\,\pi} \frac{ |{\mathbf p_2}|}{m^2_1}
|V_{cd}V_{uq}^\dagger a_1 f_{V} m_{V} |^2
\left(H_0^{B_c\to \overline B^0}(m^2_{V})\right)^2
\\
\Gamma(B_c^-\to \overline B^{\ast\,0} P^-) & = &
\frac{G^2_F}{16\,\pi} \frac{ |{\mathbf p_2}|}{m^2_1}
|V_{cd}V_{uq}^\dagger a_1 f_{P} m_{P} |^2
\left( H_t^{B_c^-\to \overline B^{\ast\,0}}(m^2_{P})\right)^2
\\
\Gamma(B_c^-\to \overline B^{\ast\,0} V^-) & = &
\frac{G^2_F}{16\,\pi} \frac{ |{\mathbf p_2}|}{m^2_1}
|V_{cd}V_{uq}^\dagger a_1 f_{V} m_{V} |^2
\sum\limits_{i=0,\pm}
\left( H_i^{B_c^-\to \overline B^{\ast\,0}}(m^2_{V})\right)^2
\\
&&\\
\Gamma(B_c^-\to \overline B_s^0 P^-) & = &
\frac{G^2_F}{16\,\pi} \frac{ |{\mathbf p_2}|}{m^2_1}
|V_{cs}V_{uq}^\dagger a_1 f_{P} m_{P} |^2
\left(H_t^{B_c^-\to \overline B_s^0}(m^2_{P})\right)^2
\\
\Gamma(B_c^-\to \overline B_s^0 V^-) & = &
\frac{G^2_F}{16\,\pi} \frac{ |{\mathbf p_2}|}{m^2_1}
|V_{cs}V_{uq}^\dagger a_1 f_{V} m_{V} |^2
\left(H_0^{B_c^-\to \overline B_s^0}(m^2_{V})\right)^2
\\
\Gamma(B_c^-\to \overline B_s^{\ast\,0} P^-) & = &
\frac{G^2_F}{16\,\pi} \frac{ |{\mathbf p_2}|}{m^2_1}
|V_{cs}V_{uq}^\dagger a_1 f_{P} m_{P} |^2
\left( H_t^{B_c^-\to \overline B_s^{\ast\,0}}(m^2_{P})\right)^2
\\
\Gamma(B_c^-\to \overline B_s^{\ast\,0} V^-) & = &
\frac{G^2_F}{16\,\pi} \frac{ |{\mathbf p_2}|}{m^2_1}
|V_{cs}V_{uq}^\dagger a_1 f_{V} m_{V} |^2
\sum\limits_{i=0,\pm}
\left( H_i^{B_c^-\to \overline B_s^{\ast\,0}}(m^2_{V})\right)^2
\\
&&\\
\Gamma(B_c^-\to B^- \pi^0) & = &
\frac{G^2_F}{16\,\pi} \frac{ |{\mathbf p_2}|}{m^2_1}
\left|V_{cd}V_{ud}^\dagger \frac{a_2}{\sqrt{2}} f_{\pi} m_{\pi}\right|^2
\left( H_t^{B_c^-\to B^-}(m^2_{\pi^0})\right)^2
\\
\Gamma(B_c^-\to B^- \rho^0) & = &
\frac{G^2_F}{16\,\pi} \frac{ |{\mathbf p_2}|}{m^2_1}
\left|
V_{cd}V_{ud}^\dagger \frac{a_2}{\sqrt{2}} f_{\rho} m_{\rho}
\right|^2
\left( H_0^{B_c^-\to B^-}(m^2_{\pi^0})\right)^2
\\
\Gamma(B_c^-\to B^-  K^0) & = &
\frac{G^2_F}{16\,\pi} \frac{ |{\mathbf p_2}|}{m^2_1}
|V_{cs}V_{ud}^\dagger a_2 f_{K} m_{K} |^2
\left( H_t^{B_c^-\to B^-}(m^2_{K^{0}})\right)^2
\\
\Gamma(B_c^-\to B^- K^{\ast\,0}) & = &
\frac{G^2_F}{16\,\pi} \frac{ |{\mathbf p_2}|}{m^2_1}
|V_{cs}V_{ud}^\dagger a_2 f_{K^{\ast}} m_{K^{\ast}}  |^2
\left( H_0^{B_c^-\to B^-}(m^2_{K^{\ast\,0}})\right)^2
\\
\Gamma(B_c^-\to B^- \overline K^0) & = &
\frac{G^2_F}{16\,\pi} \frac{ |{\mathbf p_2}|}{m^2_1}
|V_{cd}V_{us}^\dagger a_2 f_{\overline K} m_{\overline K}|^2
\left( H_t^{B_c^-\to B^-}(m^2_{\overline K^0})\right)^2
\\
\Gamma(B_c^-\to B^- \overline K^{\ast 0}) & = &
\frac{G^2_F}{16\,\pi} \frac{ |{\mathbf p_2}|}{m^2_1}
|V_{cd}V_{us}^\dagger a_2 f_{\overline K^{\ast}} m_{\overline K^{\ast}}  |^2
\left( H_0^{B_c^-\to B^-}(m^2_{\overline K^{\ast\,0}})\right)^2
\end{eqnarray*}

\end{document}